\documentclass[aps,prd,twocolumn,nofootinbib]{revtex4}

\usepackage{graphicx}
\usepackage{hyperref}
\usepackage{slashed}
\usepackage{color}

\begin{document}

\title{Heavy quarkonium wave functions at the origin and excited heavy quarkonium production via top quark decays at the LHC}

\author{Qi-Li Liao}
\email{xiaosueer@163.com}
\author{Guo-Ya Xie}
\address{College of Mobile Telecommunications Chongqing  University of Posts and Telecom, Chongqing 401520, P.R. China}

\date{\today}
\begin{abstract}
The value of the quarkonium wave function at the origin is an important quantity for studying many physical problems concerning a heavy quarkonium. This is because it is widely used to evaluate the production and decay amplitudes of the heavy quarkonium within the effective field theory framework, e.g., the nonrelativistic QCD (NRQCD). In this paper, the values of the Schr${\rm \ddot{o}}$dinger radial wave function or its first nonvanishing derivative at zero quark-antiquark separation,  i.e., $|(|c\bar{c})[n]\rangle$, $|(|b\bar{c})[n]\rangle$, and $|(b\bar{b})[n]\rangle$ quarkonium, have been tabulated under five potential models with new parameters for the heavy quarkonium. Moreover, the production of the lower-level Fock states $|(b\bar{Q})[1S]\rangle$ and $|(b\bar{Q})[1P]\rangle$, together with the higher excited Fock states $|(b\bar{Q})[nS]\rangle$ and $|(b\bar{Q})[nP]\rangle$ ($Q$ stands for the $c$ or $b$ quark; $n=2, \cdots, 6 $) through top quark decays has been studied with the new values of heavy quarkonium wave functions at the origin under the framework of NRQCD. At the LHC with the luminosity ${\cal L}\propto 10^{34}cm^{-2}s^{-1}$ and the center-of-mass energy $\sqrt{S}=14$ TeV, sizable heavy quarkonium events can be produced through top quark decays , i.e., $4 \times10^5$ $B_c$ and $B^*_c$, and $2 \times10^4$ $\eta_b$ and $\Upsilon$  events per year can be obtained according to our calculation.\\

\noindent {\bf PACS numbers:} 12.38.Bx, 14.65.Ha, 14.40.Nd, 14.40.Pq

\end{abstract}

\maketitle

\section{Introduction}

Among the heavy quarkona, the $B_c$ meson being the unique meson with two different heavy quarks in the Standard Model has aroused great interest since its discovery by the CDF collaboration \cite{cdf}. Although the ``direct" hadronic production of the $B_c$ meson has been systematically studied in Refs.~\cite{sb,skr,bc1,bc2,bc3,bcvegpy}, as a compensation to understand the production mechanism and the $B_c$ meson properties, it would be helpful to study its production ``indirectly" through $t$($\bar{t}$)-quark, $Z^0$-boson, and $W^{\pm}$-boson decays, as too many directly produced $B_c$ events shall be cut off by the trigging condition \cite{yellow,cms}. A systematical study on the indirect production of $B_c$ mesons through $t$($\bar{t}$)-quark, $Z^0$-boson,  and $W^{\pm}$-boson decay can be found in Refs.~\cite{tbc1,tbc2}, Refs.~\cite{zbc0,zbc1,zbc2,zbc3}, and Refs.~\cite{w,wbc1}, respectively. Meanwhile, it has been found that the higher excited states like $nS$ and $nP$ wave states can provide sizable contributions in the $B_c$ meson's indirect production through $W^\pm$-boson decay in Ref.~\cite{wbc2}; one should take these higher Fock states' contributions into account so as to make a better estimation with other indirect production mechanisms. Therefore, to present a systematic study on higher Fock states' indirect production of $B_c$ meson indirect production through top quark decays under the nonrelativistic quantum chromodynamics framework (NRQCD) \cite{nrqcd} is one of the purposes of the present paper.

With the LHC running at the center-of-mass energy $\sqrt{S}=14$ TeV and luminosity raising up to ${\cal L}\propto 10^{34}cm^{-2}s^{-1}$, about $10^{8}$ $t$-quark or $\bar{t}$-quark events per year will be produced \cite{bc1,bc2}. This makes the LHC much better  than Tevatron, since more $t(\bar{t})$-quark rare decays can be adopted for precise studies. A systematic study on the production of the $\bar{B}_c$ or $B_c$ meson and its excited states via $t$-quark or $\bar{t}$-quark decays, e.g. $t\rightarrow |(b\bar{c})[n]\rangle + cW^{+}$ or $\bar{t}\rightarrow |(\bar{b}c)[n]\rangle +\bar{c}W^{-}$, can be found in Refs.~\cite{tbc1,tbc2}, where $n=1S,1P$ wave states. Their results show that large number of heavy quarkonium events through top quark decays can be found at LHC(SLHC, DLHC, and TLHC, etc. \cite{ab}), so these channels shall be helpful for studying heavy quarkonium properties. In addition to the production of the two color-singlet $S$ wave states $|(b\bar{Q})(n^1S_0)\rangle$ and $|(b\bar{Q})(n^3S_1)\rangle$, naive NRQCD scaling rule shows that the production of the four color-singlet $P$-wave states $|(b\bar{Q})(n^1P_1)\rangle$ and $|(b\bar{Q})(n^3P_J)\rangle$ (with $J=0, 1, 2$; $n=1, \cdots, 6$) shall also give sizable contributions in $t\to |(b\bar{Q})[n]\rangle+W^{+}Q$ ,  where $Q$ stands for $c$ or $b$ quark accordingly. These higher excited $|(b\bar{Q})[n]\rangle$ quarkonium states, may directly or indirectly (in a cascade way) decay to its ground state with almost 100\% possibility via electromagnetic or hadronic interactions. So, it would be interesting to study higher Fock states' contributions so as to make a more sound estimation of the production of the heavy quarkonium through top quark decays, and hence to be a useful reference for experimental studies.

In the framework of an effective theory of the NRQCD, a doubly heavy meson system is considered as an expansion of various Fock states. The relative importance among those infinite ingredients is evaluated by the velocity scaling rule. In evaluating the production and decay amplitude of the heavy quarkonium, each factor can be separated into a short-distance factor and a long-distance coefficient. The short-distance factor can be computed using perturbative quantum chromodynamics (pQCD), and the long-distance factor is associated with quarkonia structure, which is expressed in terms of nonperturbative matrix elements $\langle{\cal O}^H(n)\rangle$. The matrix elements $\langle{\cal O}^H(n) \rangle$ can be expressed in terms of the meson's nonrelativistic wave function, or its derivatives, evaluated at the origin under the color-singlet model \cite{creh}, and the wave function is identified with the Schr${\rm \ddot{o}}$dinger wave function calculated in potential models for heavy quarkonium. As a result, the rigorously calculated nonrelativistic wave function at the origin is very important in studying heavy quarkonium decay and production.

Because of the emergence of massive fermion lines in $t\to |(b\bar{Q})[n]\rangle+W^{+}Q$, the analytical expression for the squared amplitude becomes too complex and lengthy. For such complicated processes, one important way out is to deal with it directly at the amplitude level. For this purpose, the ``improved trace technology" suggested and developed by Refs.~\cite{sb,tbc2,zbc0,zbc1,zbc2} shows that the hard scattering amplitude can also be expressed by the dot products of the concerned particle momenta like that of the squared amplitudes. In the present paper, we shall adopt improved trace technology to derive the analytical expression for all the mentioned Fock states, and to be a useful reference, we shall transform its form to be as compact as possible by fully applying symmetries and relations among them.

This paper is organized as follows: In Sec. II, we show our calculation techniques for the mentioned top quark semi exclusive decays to heavy quarkonium. In order to calculate the production of the excited heavy quarkonium via top quark decays, we present five QCD-motivated potential models for heavy $|(Q\bar{Q'})[n]\rangle$ quarkonium in Sec.III. In Sec.IV, we calculate and tabulate all the values of the Schr${\rm \ddot{o}}$dinger radial wave functions, its first nonvanishing derivative, and its second nonvanishing derivative at zero quark-antiquark separation. Furthermore, we also present numerical results and make some discussions on the properties of the heavy quarkonium production through top quark decays. The final section is reserved for a summary.

\section{CALCULATION TECHNIQUES}

We shall deal with some typical top quark semiexclusive processes for the heavy quarkonium production, i.e., $t (q_{0}) \to |(b\bar{Q})[n]\rangle(q_3) +W^{+}(q_2) + Q(q_1)$, where $q_i$ ($i=0, 1, 2, 3$) are the momenta of the corresponding particles. According to the NRQCD factorization formula \cite{nrqcd}, its total decay widths $d\Gamma$ can be factorized as
\begin{equation}
d\Gamma=\sum_{n} d\hat\Gamma(t\to |(b\bar{Q})[n]\rangle+ Q+W^{+}) \langle{\cal O}^H(n) \rangle,
\end{equation}
where the nonperturbative matrix element $\langle{\cal O}^{H}(n)\rangle$ describes the hadronization of a $b\bar{Q}$ pair into the observable quark state $H$ and is proportional to the transition probability of the perturbative state $b\bar{Q}$ into the bound state $|(b\bar{Q})[n]\rangle$. As for the color-singlet components, the matrix elements can be directly related to the Schr${\rm \ddot{o}}$dinger wave functions at the origin for the $S$ wave states, the first derivative of the wave functions at the origin for the $P$ wave states, or the second derivative of the wave functions at the origin for the $D$ wave states \cite{nrqcd}, which can be computed via potential NRQCD (pNRQCD) \cite{pnrqcd1,yellow}, lattice QCD \cite{lat1} and/or the potential models \cite{pot1,pot2,pot3,pot4,pot5}.

The short-distance decay widths are given by
\begin{equation}
d\hat\Gamma(t\to |(b\bar{Q})[n]\rangle+Q+W^{+})= \frac{1}{2 q^0_{0}} \overline{\sum}  |M|^{2} d\Phi_3,
\end{equation}
where $\overline{\sum}$ means that we need to average over the spin and color states of the initial particle and to sum over the color and spin of all the final particles.

In the top quark rest frame, the three-particle phase space can be written as
\begin{equation}
d{\Phi_3}=(2\pi)^4 \delta^{4}\left(q_{0} - \sum_f^3 q_{f}\right)\prod_{f=1}^3 \frac{d^3{\vec{q}_f}}{(2\pi)^3 2q_f^0}.
\end{equation}

We have done a calculation to simplify the $1 \to 3$ phase space with massive quark/antiqark in the final state in Refs.~\cite{BK,tbc2,zbc1}. To shorten the paper, we shall not present it here and the interested reader may turn to these three references for the detailed technology. With the help of the formulas listed in Refs.~\cite{BK,tbc2,zbc1}, one can not only derive the whole decay widths but also obtain the corresponding differential decay widths that are helpful for experimental studies, such as $d\Gamma/ds_{1}$, $d\Gamma/ds_{2}$, $d\Gamma/d\cos\theta_{12}$, and $d\Gamma/d\cos\theta_{13}$, where $s_{1}=(q_1+q_2)^2$, $s_{2}=(q_1+q_3)^2$, $\theta_{12}$ is the angle between $\vec{q}_1$ and $\vec{q}_2$, and $\theta_{13}$ is the angle between $\vec{q}_1$ and $\vec{q}_3$.

In particular, the partial decay widths over $s_{1}$ and $s_{2}$ can be expressed as
\begin{equation}
d\Gamma= \frac{\langle{\cal O}^H(n) \rangle}{256 \pi^3 m^3_t}( \overline{\sum}|M|^{2}) ds_{1}ds_{2}.
\end{equation}
where $m_t$ is the mass of the top quark.

The color-singlet nonperturbative matrix element $\langle{\cal O}^H(n) \rangle$ can be related to the Schr${\rm \ddot{o}}$dinger wave function $\psi_{(b\bar{Q})}(0)$ at the origin or the first derivative of the wave function $\psi^\prime_{(b\bar{Q})}(0)$ at the origin for $S$- and $P$-wave quarkonium states. For convenience, we have adopted the convention of Refs.~\cite{nrqcd, yellow} for the nonperturbative matrix element.
\begin{eqnarray}
\langle{\cal O}^H(nS) \rangle &\simeq& |\psi_{\mid(b\bar{Q})[nS]\rangle}(0)|^2,\nonumber\\
\langle{\cal O}^H(nP) \rangle &\simeq& |\psi^\prime_{\mid(b\bar{Q})[nP]\rangle}(0)|^2.
\end{eqnarray}
Since the spin-splitting effects are small, we will not distinguish the difference between the wave function parameters for the spin-singlet and spin-triplet states at the same $n$th level.

And then our task is to deal with the hard-scattering amplitude for specified processes
\begin{eqnarray}
&& t\rightarrow |(b\bar{c})[n]\rangle + cW^{+},~~t\rightarrow |(b\bar{b})[n]\rangle + bW^{+}.
\end{eqnarray}

\begin{figure}
\includegraphics[width=0.45\textwidth]{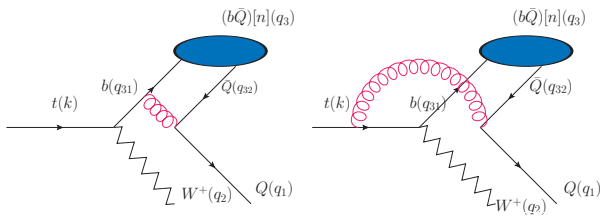}
\caption{Feynman diagrams for the process $t(q_{0})\rightarrow |(b\bar{Q})[n]\rangle(q_3) + W^{+}(q_2)+ Q(q_1)$, where $Q$ stands for the $c$ and $b$ quark in the left and right panels, respectively. $|(b\bar{Q})[n]\rangle$ quarkonium stands for a heavy quarkonium Fock state.} \label{feyn1}
\end{figure}

For convenience, we shorten the two processes as $t(q_{0}) \rightarrow |(b\bar{Q})[n]\rangle(q_3) + W^{+}(q_2)+Q(q_1)$, where $Q$ stands for $c$ or $b$ quark accordingly. The Feynman diagrams of the process are presented in Fig.~\ref{feyn1}, where the intermediate gluon should be hard enough to produce a $c\bar{c}$ pair or $b\bar{b}$ pair, so the amplitude is pQCD calculable.

These amplitudes can be generally expressed as
\begin{equation} \label{amplitude}
iM = {\cal{C}} {\bar {u}_{s i}}({q_1}) \sum\limits_{n = 1}^{m} {{\cal A} _n } {u_{s' j}}({k}),
\end{equation}
where $m$ stands for the number of Feynman diagrams, $s$ and $s'$ are spin states, and $i$ and $j$ are color indices for the outing $Q$ quark and the initial top quark, respectively. The overall factor ${\cal C}=\frac{2gg_s^2 V_{tb}}{\sqrt{6}}\delta_{ij}$, here $V_{tb}$ is the Cabibbo-Kobayashi-Maskawa (CKM) matrix element. ${\cal A} _n $`s in the formulas are listed in Ref.~\cite{tbc2}.

As mentioned above, we adopt the improved trace technology to simplify the amplitudes $M_{ss^{\prime}}$ at the amplitude level. In a difference from the helicity amplitude approach \cite{bcvegpy,helicity1,helicity2,helicity3}, only the coefficients of the basic Lorentz structures are numerical at the amplitude level. However, by using the improved trace technology in Refs.~\cite{tbc2,zbc0,zbc1,zbc2,wbc1,wbc2}, one can sequentially obtain the squared amplitudes, and the numerical efficiency can also be greatly improved. The standard procedures of the improved trace technology for $t(q_{0}) \rightarrow |(b\bar{Q})[n]\rangle(q_3) + W^{+}(q_2)+Q(q_1)$ have been presented in Ref.~\cite{tbc2}.

\section{Potential model}

Nonperturbative matrix elements  $\langle{\cal O}^H(n) \rangle$ can be related to the wave function at the origin \cite{nrqcd}. In the rest frame of the $|(Q\bar{Q'})[n]\rangle$ quarkonium, it is convenient to separate the Schr${\rm \ddot{o}}$dinger wave function into radial and angular pieces as
\begin{eqnarray}
\Psi_{nlm}(\vec{r})&=&R_{nl}(r)Y_{lm}(\theta,\varphi),
\end{eqnarray}
where $n$ is the principal quantum number, and $l$ and $m$ are the orbital angular momentum quantum number and its projection. $R_{nl}(r)$ and $Y_{lm} (\theta,\varphi)$ are the radial wave function and the spherical harmonic function accordingly.

Further on, the value of the radial wave function, its first nonvanishing derivative or its second nonvanishing derivative at the origin can be obtained as in \cite{nrqcd}
\begin{eqnarray}
R^{(l)}_{nl}(0)&=&\frac{d^{l}R_{nl}(r)}{{dr}^l}\mid_{r=0},
\end{eqnarray}
where $l=0$, $l=1$, and $l=2$ correspond to the radial wave functions $R_{|(Q\bar{Q}')[nS]\rangle}(0)$,  $R^{'}_{|(Q\bar{Q}')[nP]\rangle}(0)$,  and $R^{''}_{|(Q\bar{Q}')[nD]\rangle}(0)$ at the origin.

The wave function $\Psi_{|(Q\bar{Q}')[nS]\rangle}(0)$, the first derivative of the wave function $\Psi^{'}_{|(Q\bar{Q}')[nP]\rangle}(0)$, and the second derivative of the wave function $\Psi^{''}_{|(Q\bar{Q}')[nD]\rangle}(0)$ at the origin are related to the radial wave function $R_{|(Q\bar{Q}')[nS]\rangle}(0)$, the first derivative of the radial wave function $R^{'}_{|(Q\bar{Q}')[nP]\rangle}(0)$, and the second derivative of the radial wave function $R^{''}_{|(Q\bar{Q}')[nD]\rangle}(0)$ at the origin, accordingly.
\begin{eqnarray}
\Psi_{|(Q\bar{Q'})[nS]\rangle}(0)&=&\sqrt{{1}/{4\pi}}R_{|(Q\bar{Q}')[nS]\rangle}(0),\nonumber\\
\Psi'_{|(Q\bar{Q'})[nP]\rangle}(0)&=&\sqrt{{3}/{4\pi}}R'_{|(Q\bar{Q}')[nP]\rangle}(0),\nonumber\\
\Psi{''}_{|(Q\bar{Q'})[nD]\rangle}(0)&=&\sqrt{{5}/{16\pi}}R{''}_{|(Q\bar{Q}')[nD]\rangle}(0).
\end{eqnarray}

Next, we will give a brief introduction to the five QCD-motivated potentials that give reasonable accounts of the $|(c\bar{c})[n]\rangle$, $|(b\bar{c})[n]\rangle$, and $|(b\bar{b})[n]\rangle$ quarkonium.

(1) Buchm${\rm \ddot{u}}$ller and Tye have given the QCD-motivated potential (B.T. potential) with two-loop correction \cite{pot2,wgs,ec} as
\begin{eqnarray}
V(r)= k r-\frac{8 \pi}{3 \beta_0} \frac{v(\lambda~{r})}{r},~~~~r\geq0.01fm,
\end{eqnarray}
and
\begin{widetext}
\begin{eqnarray}
V(r)&=&-\frac{16 \pi}{3 \beta_0} \frac{1}{r\ln(1/\Lambda^2_{\overline{MS}}~r^2)}\left[1+(r_{E}+\frac{53}{3 \beta_0})\frac{1}{r\ln(1/\Lambda^2_{\overline{MS}}~r^2)}-\frac{\beta_1}{\beta^2_0}\frac{\ln\ln(1/\Lambda^2_{\overline{MS}}~r^2)}{\ln(1/\Lambda^2_{\overline{MS}}~r^2)}\right],   ~~~~r<0.01fm.
\end{eqnarray}
\end{widetext}
in which $k=\frac{1}{2\pi\alpha'}$, with $\alpha'=1.067GeV^{-2}$, is the Regge slope. $\beta_0=11-\frac{2}{3}n_{f}$, $\beta_1=102-\frac{38}{3}n_f$, where here $n_f$ is the number of active flavor quarks. ${\Lambda_{\overline{MS}}}$ stands for the scale parameters, and $\overline{MS}$ is the modified minimal subtraction scheme. The  parameter $\lambda=(\frac{3 \beta_0 k}{8\pi})^{1/2}$ can be expressed in terms of the string constant $k$.
And
\begin{displaymath}
v(r)=\frac{4 \beta_0}{\pi}\int^\infty_0\frac{dq}{q}\left[\hat{\rho}(q^2)-\frac{K}{q^2}\right]\sin(\frac{q}{\lambda}~r),
\end{displaymath}
with
\begin{displaymath}
K=\frac{\Lambda^2_{\overline{MS}}}{\beta_0}\exp{\left[-\frac{\beta_1}{\beta^2_0}(\gamma_E+\ln\frac{l}{\beta_0})\right]},
\end{displaymath}
where $l=24$, and $\gamma_E=0.5772$ is the Euler constant.
$\hat{\rho}(q^2)$ is a physical quantity and therefore independent of the choice of gauge and the subtraction scheme. For small values of $q^2$, it has the form of
\begin{displaymath}
\hat{\rho}(q^2)~~{\overrightarrow{q^2\rightarrow{0}}}~~\frac{K}{q^2}.
\end{displaymath}
for large $q^2$, perturbative QCD implies
\begin{eqnarray}
\hat{\rho}(q^2)~~&{\overrightarrow{q^2\rightarrow {\infty}}}&~ \frac{1}{\beta_0\ln(q^2/\Lambda^2_{\overline{MS}})}-\frac{\beta_1}{\beta^3_0}\frac{\ln\ln(q^2/\Lambda^2_{\overline{MS}})}{\ln^2(q^2/\Lambda^2_{\overline{MS}})}\nonumber\\
&&+O(\frac{1}{\ln^3(q^2/\Lambda^2_{\overline{MS}})}).
\end{eqnarray}
Here $q$ is the transfer momentum in the rest frame of the $|(c\bar{c})[n]\rangle$ , $|(b\bar{c})[n]\rangle$ , and $|(b\bar{b})[n]\rangle$ quarkonium.

(2) The QCD-motivated potential with one-loop correction is given by John L. Richardson (J. potential) \cite{jlr} as
\begin{eqnarray}
V(r)=\frac{8 \pi}{3 \beta_0} \Lambda_{\overline{MS}}\left[\Lambda_{\overline{MS}}~r-\frac{f({\Lambda_{\overline{MS}}~r)}}{\Lambda_{\overline{MS}}~ r}\right],
\end{eqnarray}
where
\begin{displaymath}
f(t)=\frac{4}{\pi} \int^{\infty}_{0}dq\frac{\sin(qt)}{q}\left[\frac{1}{\ln{(1+q^2)}}-\frac{1}{q^2}\right].
\end{displaymath}

(3) The QCD-motivated potential with two-loop correction is given by K. Igi and S. Ono (I.O. potential) \cite{kso,sr} as
\begin{eqnarray}
V(r)&=&-\frac{4\alpha_s(\mu)}{3 r} +\left[1+\frac{\alpha_s(\mu)}{2\pi}(\beta_0\ln\mu r+A)\right],
\end{eqnarray}
with
\begin{displaymath}
\alpha_s(\mu)=\frac{4\pi}{\beta_0\ln(\mu^2/\Lambda^2_{\overline{MS}})} \left[1-\frac{\beta_1}{\beta^2_0}\frac{\ln\ln(\mu^2/\Lambda^2_{\overline{MS}})}{\ln(\mu^2/\Lambda^2_{\overline{MS}})}\right],
\end{displaymath}
in which $A=\beta_0 \gamma_E+\frac{31}{6}-\frac{5}{9}n_f$, $\mu=\frac{m_Q m_Q'}{m_Q+ m_Q'}$, where $n_{f}$ has the same meaning of the first potential model, and $m_{Q}$ or $m_{Q'}$ is the mass of the heavy quark $Q$ or $Q'$, accordingly.

(4) The QCD-motivated potential with two loop correction is given by Yu-Qi Chen and Yu-Ping Kuang (C.K. potential) as \cite{pot5,sr}
\begin{eqnarray}
V(r)&=&k r -\frac{16 \pi}{3 \beta_0} \frac{1}{ r f(r)} \left[1-\frac{\beta_1}{\beta^2_0}\frac{\ln{f(r)}}{f(r)}+\frac{2 \gamma_E+\frac{93-10 n_f}{9 \beta_0}}{f(r)}\right],
\end{eqnarray}
with
\begin{displaymath}
f(r)=\ln^2(\frac{1GeV}{\Lambda_{\overline{MS}}}+4.62-B)
\end{displaymath}
and
\begin{displaymath}
B(r)=(1-\frac{\Lambda_{\overline{MS}}}{4\Lambda^I_{\overline{MS}}})\frac{1-\exp\left[-(15(3\frac{\Lambda^I_{\overline{MS}}}{\Lambda_{\overline{MS}}}-1)\Lambda_{\overline{MS}}~r)^2)\right]}{\Lambda_{\overline{MS}}~ r},
\end{displaymath}
where $\Lambda^I_{\overline{MS}}=0.18GeV$.

(5) The QCD-motivated Coulomb-plus-linear potential (Cor. potential) \cite{pot1,ec,sr} has the form of
\begin{eqnarray}
V(r)&=&-\frac{0.47}{r}+0.19{GeV}^2\times r+0.051GeV.
\end{eqnarray}

\section{Numerical Results}

\subsection{Input parameters}

For calculating the wave function at the origin of the five potentials \cite{phfp}, we adopt the scale parameters ${\Lambda_{\overline{MS}}}$ as ${\Lambda^{n_f=3}_{\overline{MS}}}$=0.386 GeV, ${\Lambda^{n_f=4}_{\overline{MS}}}$=0.332 GeV, ${\Lambda^{n_f=5}_{\overline{MS}}}$=0.231 GeV, ${\Lambda^{n_f=6}_{\overline{MS}}}$=0.0938 GeV \cite{pdg}. The quark mass is adopted as the values of the constituent quark mass of the $|(Q\bar{Q}')[n]\rangle$ quarkonium derived in Refs.~\cite{pdg,pot5,sr,tse}. The quantities $|R_{|(Q\bar{Q}')[nS]\rangle}(0)|^2$, $|R^{'}_{|(Q\bar{Q}')[nP]\rangle}(0)|^2$, and $|R^{''}_{|(Q\bar{Q}')[nD]\rangle}(0)|^2$ are presented in Tables \ref{tabrpa}, \ref{tabrpb}, and \ref{tabrpc} for the five potential models. During the following calculation, we adopt the values of wave functions at the origin under the B.T. potential  as the central values for calculations of the decay widths of $t\rightarrow |(b\bar{Q})[n]\rangle +QW^{+}$, [$n_{f}$=3 is for $|(b\bar{c})[n]\rangle$ quarkonium and $n_{f}$=4 for $|(b\bar{b})[n]\rangle$ quarkonium], since it is noted that the B.T. model potential has the correction of two-loop short-distance behavior in pQCD \cite{pot2}. The results for the other four potential models, i.e., the J. model \cite{jlr}, the I.O. model \cite{kso}, the C.K. model \cite{pot5}, and the Cor. model \cite{pot1}, will be adopted as an error analysis.
\begin{widetext}
\begin{center}
\begin{table}
\caption{Radial wave functions at the origin and related quantities for $|(c\bar{c})[n]\rangle$ quarkonium.}
\begin{tabular}{|c||c|c|c|c|c|c|c|c}
\hline
$~~~~~~|(c\bar{c})[n]\rangle~~~~~~$ &~~~Mass and potential~~~&~~~$n=1$~~~&~~~$n=2$~~~&~~~$n=3$~~~&~~~$n=4$~~~&~~~$n=5$~~~&~~~$n=6$~~~\\
\hline
~~~& $m_{c}~({GeV})$&1.48&1.82&1.92&2.02&2.12&2.25\\
~~~~~&B.T.($n_{f}$=3) \cite{pot2}&2.458&1.617&0.969&0.796&0.701&0.721\\
~~~~~&B.T.($n_{f}$=4) \cite{pot2}&2.344&1.360&0.882&0.793&0.747&0.722\\
$S$ states&J.~~~~($n_{f}$=3) \cite{jlr}&1.119&1.057&0.985&0.970&0.976&1.008\\
~~~~&J.~~~~($n_{f}$=4) \cite{jlr}&0.997&0.910&0.836&0.816&0.817&0.841\\
$|R_{|[nS]\rangle}(0)|^2$&I.O.~($n_{f}$=3) \cite{kso}&0.565&0.549&0.518&0.513&0.519&0.538\\
$({GeV}^3)$&I.O.~($n_{f}$=4) \cite{kso}&0.599&0.570&0.534&0.527&0.532&0.551\\
~~~&C.K.($n_{f}$=3) \cite{pot5}&0.726&0.614&0.558&0.542&0.541&0.557\\
~~~~&C.K.($n_{f}$=4) \cite{pot5}&0.795&0.652&0.584&0.564&0.560&0.574\\
~~~~&Cor.~~~~~~~~~~ \cite{pot1}&0.974&0.889&0.821&0.807&0.812&0.842\\
\hline
~~~& $m_{c}~({GeV})$&1.75&1.96&2.12&2.26&2.38&~~~\\
~~~&B.T.($n_{f}$=3) \cite{pot2}&0.322&0.224&0.387&0.467&0.499&~~~\\
~~~&B.T.($n_{f}$=4) \cite{pot2}&0.329&0.230&0.378&0.474&0.514&~~~~\\
$P$ states&J.~~~~($n_{f}$=3) \cite{jlr}&0.172&0.309&0.437&0.566&0.694&~~~\\
~~~~~&J.~~~~($n_{f}$=4) \cite{jlr}&0.135&0.237&0.332&0.427&0.521&~~~~\\
$|R^{'}_{|[nP]\rangle}(0)|^2$&I.O.~($n_{f}$=3) \cite{kso} &0.053&0.099&0.142&0.186&0.231&~~~\\
$({GeV}^5)$&I.O.~($n_{f}$=4) \cite{kso} &0.057&0.104&0.149&0.195&0.240&~~~~\\
~~~~&C.K.($n_{f}$=3) \cite{pot5}&0.074&0.128&0.177&0.226&0.275&~~~\\
~~~~~~~~&C.K.($n_{f}$=4) \cite{pot5}&0.081&0.139&0.191&0.243&0.294&~~~~\\
~~~~~& Cor.~~~~~~~~~~ \cite{pot1} &0.091&0.169&0.244&0.320&0.376&~~~\\
\hline
~~~& $m_{c}~({GeV})$&1.88&2.07&2.23&2.36&~~~&~~~\\
~~~& B.T.($n_{f}$=3) \cite{pot2}&0.033&0.218&0.377&0.502&~~~&~~~\\
~~~& B.T.($n_{f}$=4) \cite{pot2}&0.048&0.203&0.359&0.521&~~~~&~~~~\\
$D$ states&J.~~~~($n_{f}$=3) \cite{jlr}&0.099&0.274&0.521&0.830&~~~&~~~\\
~~~~&J.~~~~($n_{f}$=4) \cite{jlr}&0.066&0.181&0.342&0.542&~~~~&~~~~\\
$|R^{''}_{|[nD]\rangle}(0)|^2$&I.O.~($n_{f}$=3) \cite{kso} &0.020&0.056&0.108&0.175&~~~&~~~\\
$({GeV}^7)$&I.O.~($n_{f}$=4) \cite{kso}&0.021&0.059&0.115&0.185&~~~~&~~~~\\
~~~& C.K.($n_{f}$=3) \cite{pot5}&0.028&0.076&0.144&0.228&~~~&~~~\\
~~~~& C.K.($n_{f}$=4) \cite{pot5}&0.030&0.083&0.156&0.246&~~~~&~~~~\\
~~~~& Cor.~~~~~~~~~~ \cite{pot1}&0.036&0.104&0.202&0.328&~~~&~~~\\
\hline
\end{tabular}
\label{tabrpa}
\end{table}
\end{center}

\begin{center}
\begin{table}
\caption{Radial wave functions at the origin and related quantities for $|(b\bar{c})[n]\rangle$ quarkonium.}
\begin{tabular}{|c||c|c|c|c|c|c|c|c}
\hline
$~~~~~~|(b\bar{c})[n]\rangle~~~~~~$ &~~~Mass and potential~~~&~~~$n=1$~~~&~~~$n=2$~~~&~~~$n=3$~~~&~~~$n=4$~~~&~~~$n=5$~~~\\
\hline
~~~& $m_{c}~({GeV})$&1.45&1.82&1.96&2.10&2.15\\
~~~&$m_{b}~({GeV})$&4.85&5.03&5.15&5.30&5.45\\
~~~~~&B.T.($n_{f}$=3) \cite{pot2}&3.848&1.987&1.347&1.279&1.118\\
~~~~~&B.T.($n_{f}$=4) \cite{pot2}&4.009&1.397&1.209&1.295&1.218\\
~~~~~&B.T.($n_{f}$=5) \cite{pot2}&3.600&2.478&1.405&1.074&1.132\\
$S$ states&J.~~~~($n_{f}$=3) \cite{jlr}&2.021&1.805&1.656&1.623&1.571\\
~~~~~~~&J.~~~~($n_{f}$=4) \cite{jlr}&1.829&1.567&1.414&1.372&1.319\\
~~~~~~~&J.~~~~($n_{f}$=5) \cite{jlr}&1.331&1.050&0.915&0.872&0.828\\
$|R_{|[nS]\rangle}(0)|^2$&I.O.~($n_{f}$=3) \cite{kso}&6.211&2.169&1.301&0.941&0.734\\
$({GeV}^3)$&I.O.~($n_{f}$=4) \cite{kso}&5.262&1.958&1.186&0.865&0.677\\
~~~~~~&I.O.~($n_{f}$=5) \cite{kso}&3.584&1.477&0.914&0.678&0.534\\
~~~~&C.K.($n_{f}$=3) \cite{pot5}&1.304&1.046&0.933&0.903&0.868\\
~~~~&C.K.($n_{f}$=4) \cite{pot5}&1.447&1.115&0.979&0.939&0.897\\
~~~~&C.K.($n_{f}$=5) \cite{pot5}&1.636&1.202&1.034&0.982&0.932\\
~~~~&Cor.~~~~~~~~~~ \cite{pot1}&1.783&1.594&1.464&1.442&1.393\\
\hline
~~~& $m_{c}~({GeV})$&1.75&1.96&2.15&2.26&~~~\\
~~~&$m_{b}~({GeV})$&4.93&5.13&5.25&5.37&~~~\\
~~~&B.T.($n_{f}$=3) \cite{pot2}&0.518&0.500&0.729&0.823&~~~~\\
~~~&B.T.($n_{f}$=4) \cite{pot2}&0.756&0.436&0.775&0.929&~~~~\\
~~~&B.T.($n_{f}$=5) \cite{pot2}&0.895&0.930&0.745&0.862&~~~~\\
$P$ states&J.~~~~($n_{f}$=3) \cite{jlr}&0.413&0.686&0.943&1.154&~~~~\\
~~~~~~&J.~~~~($n_{f}$=4) \cite{jlr}&0.331&0.537&0.729&0.884&~~~~\\
~~~~~~&J.~~~~($n_{f}$=5) \cite{jlr}&0.160&0.246&0.325&0.387&~~~~\\
$|R^{'}_{|[nP]\rangle}(0)|^2$&I.O.~($n_{f}$=3) \cite{kso} &0.573&0.483&0.416&0.364&~~~\\
$({GeV}^5)$&I.O.~($n_{f}$=4) \cite{kso} &0.471&0.410&0.359&0.317&~~~~\\
~~~~&I.O.~($n_{f}$=5) \cite{kso} &0.289&0.265&0.241&0.216&~~~~\\
~~~~~~~~&C.K.($n_{f}$=3) \cite{pot5}&0.186&0.312&0.390&0.499&~~~~\\
~~~~~~~~&C.K.($n_{f}$=4) \cite{pot5}&0.209&0.346&0.426&0.543&~~~~\\
~~~~~~~~&C.K.($n_{f}$=5) \cite{pot5}&0.241&0.390&0.475&0.601&~~~~\\
~~~~~& Cor.~~~~~~~~~~ \cite{pot1} &0.219&0.380&0.537&0.668&~~~~\\
\hline
~~~& $m_{c}~({GeV})$&1.88&2.10&2.25&~~~&~~~\\
~~~&$m_{b}~({GeV})$&5.12&5.25&5.35&~~&~~~\\
~~~& B.T.($n_{f}$=3) \cite{pot2}&0.069&0.411&0.741&~~~~&~~~~\\
~~~& B.T.($n_{f}$=4) \cite{pot2}&0.146&0.514&0.873&~~~~&~~~~\\
~~~& B.T.($n_{f}$=5) \cite{pot2}&0.624&0.803&0.994&~~~~&~~~~\\
$D$ states&J.~~~~($n_{f}$=3) \cite{jlr}&0.299&0.789&1.400&~~~&~~~~\\
~~~~&J.~~~~($n_{f}$=4) \cite{jlr}&0.205&0.533&0.936&~~~~&~~~~\\
~~~~&J.~~~~($n_{f}$=5) \cite{jlr}&0.066&0.166&0.285&~~~~&~~~~\\
$|R^{''}_{|[nD]\rangle}(0)|^2$&I.O.~($n_{f}$=3) \cite{kso} &0.172&0.237&0.264&~~~&~~~~\\
$({GeV}^7)$&I.O.~($n_{f}$=4) \cite{kso} &0.133&0.189&0.215&~~~~&~~~~\\
~~~&I.O.~($n_{f}$=5) \cite{kso} &0.071&0.107&0.125&~~~~&~~~~\\
~~~~& C.K.($n_{f}$=3) \cite{pot5}&0.089&0.230&0.403&~~~&~~~~\\
~~~~& C.K.($n_{f}$=4) \cite{pot5} &0.100&0.256&0.444&~~~~&~~~~\\
~~~~& C.K.($n_{f}$=5) \cite{pot5} &0.115&0.290&0.500&~~~~&~~~~\\
~~~~~~& Cor.~~~~~~~~~~ \cite{pot1}&0.111&0.306&0.559&~~~&~~~~\\
\hline
\end{tabular}
\label{tabrpb}
\end{table}
\end{center}

\begin{center}
\begin{table}
\caption{Radial wave functions at the origin and related quantities for $|(b\bar{b})[n]\rangle$ quarkonium.}
\begin{tabular}{|c||c|c|c|c|c|c|c|c|c}
\hline
$~~~~~~~|(b\bar{b})[n]\rangle$~~~~~ &~~~Mass and potential~~~&~~$n=1$~~&~~$n=2$~~~&~~$n=3$~~~~&~~$n=4$~~~~&~~$n=5$~~~&~~$n=6$~~~&~~~$n=7$~~\\
\hline
~~&$m_{b}~({GeV})$&4.71&5.01&5.17&5.27&5.41&5.50&5.58\\
~~~&B.T.($n_{f}$=4) \cite{pot2}&16.12&6.746&2.172&2.588&2.665&2.576&2.377\\
~~~&B.T.($n_{f}$=5) \cite{pot2}&14.00&7.418&4.835&2.960&2.231&2.247&2.310\\
~~~&B.T.($n_{f}$=6) \cite{pot2}&8.447&4.657&3.689&3.197&2.928&2.716&2.530\\
$S$ states&J.~~~~($n_{f}$=4) \cite{jlr}&7.114&4.146&3.401&3.047&2.886&2.762&2.676\\
~~~~~~&J.~~~~($n_{f}$=5) \cite{jlr}&5.590&2.888&2.258&1.971&1.838&1.739&1.670\\
~~~~~~&J.~~~~($n_{f}$=6) \cite{jlr}&3.071&1.210&0.833&0.679&0.607&0.557&0.523\\
$|R_{|[nS]\rangle}(0)|^2$&I.O.~($n_{f}$=4) \cite{kso}&9.981&3.462&2.051&1.454&1.143&0.941&0.802\\
$({GeV}^3)$&I.O.~($n_{f}$=5) \cite{kso}&8.699&3.015&1.787&1.267&0.998&0.822&0.701\\
~~~~&I.O.~($n_{f}$=6) \cite{kso}&5.878&2.084&1.246&0.889&0.704&0.582&0.498\\
~~~&C.K.($n_{f}$=4) \cite{pot5}&5.298&2.783&2.220&1.972&1.861&1.780&1.724\\
~~~&C.K.($n_{f}$=5) \cite{pot5}&6.081&2.992&2.325&2.037&1.905&1.810&1.745\\
~~~&C.K.($n_{f}$=6) \cite{pot5}&6.823&3.151&2.380&2.055&1.904&1.797&1.724\\
~~~~&Cor.~~~~~~~~~~ \cite{pot1}&9.140&4.771&3.901&3.499&3.324&3.183&3.084\\
\hline
~~&$m_{b}~({GeV})$&4.94&5.12&5.20&5.37&5.47&5.56&~~ \\
~~~~~&B.T.($n_{f}$=4) \cite{pot2} &5.874 &2.827&2.578&3.217&3.573&3.669&~~ \\
~~~~~&B.T.($n_{f}$=5) \cite{pot2} &4.973&5.216&4.015&3.026&3.172&3.541&~~~~~\\
~~~~~&B.T.($n_{f}$=6) \cite{pot2} &1.964&2.460&2.698&3.002&3.181&3.324&~~~~~\\
$P$ states&J.~~~~($n_{f}$=4) \cite{jlr}&1.644&2.146&2.453&2.841&3.143&3.431&~~\\
~~~&J.~~~~($n_{f}$=5) \cite{jlr}&0.883&1.070&1.172&1.323&1.436&1.544&~~~~~\\
~~~&J.~~~~($n_{f}$=6) \cite{jlr}&0.205&0.206&0.201&0.212&0.219&0.226&~~~~~\\
$|R^{'}_{|[nP]\rangle}(0)|^2$&I.O.~($n_{f}$=4) \cite{kso} &1.165&0.965&0.794&0.700&0.622&0.565&~~\\
$({GeV}^5)$&I.O.~($n_{f}$=5) \cite{kso} &0.914&0.759&0.625&0.554&0.493&0.449&~~~~~\\
~~~~~~~&I.O.~($n_{f}$=6) \cite{kso} &0.496&0.418&0.346&0.310&0.278&0.254&~~~~~\\
~~~~&C.K.($n_{f}$=4) \cite{pot5}&1.111&1.324&1.450&1.636&1.778&1.915&~~\\
~~~~&C.K.($n_{f}$=5) \cite{pot5}&1.344&1.547&1.662&1.854&1.997&2.136&~~~~~\\
~~~~&C.K.($n_{f}$=6) \cite{pot5}&1.661&1.829&1.917&2.107&2.245&2.381&~~~~~\\
~~~~~& Cor.~~~~~~~~~~ \cite{pot1}&1.218&1.667&1.961&2.325&2.613&2.886&~~ \\
\hline
~~&$m_{b}~({GeV})$&5.03&5.20&5.33&5.44&5.52&~~~&~~ \\
~~~& B.T.($n_{f}$=4) \cite{pot2}&4.469&2.733&5.181&7.108&8.543&~~~&~~ \\
~~~& B.T.($n_{f}$=5) \cite{pot2}&5.621&8.007&7.114&7.327&9.038&~~~~~&~~~~~\\
~~~& B.T.($n_{f}$=6) \cite{pot2}&1.631&3.274&4.855&6.364&7.717&~~~~~&~~~~~\\
$D$ states&J.~~~~($n_{f}$=4) \cite{jlr}&1.378&2.891&4.495&6.197&8.144&~~~&~~ \\
~~~~&J.~~~~($n_{f}$=5) \cite{jlr}&0.491&0.979&1.472&1.980&2.550&~~~~~&~~~~~\\
~~~~&J.~~~~($n_{f}$=6) \cite{jlr}&0.043&0.075&0.102&0.129&0.157&~~~~~&~~~~~\\
$|R^{''}_{|[nD]\rangle}(0)|^2$&I.O.~($n_{f}$=4) \cite{kso}&0.421&0.565&0.595&0.638&0.639&~~~&~~ \\
$({GeV}^7)$&I.O.~($n_{f}$=5) \cite{kso}&0.296&0.400&0.424&0.456&0.457&~~~~~&~~~~~\\
~~~~&I.O.~($n_{f}$=6) \cite{kso}&0.129&0.177&0.190&0.205&0.207&~~~~~&~~~~~\\
~~~& C.K.($n_{f}$=4) \cite{pot5} &0.724&1.452&2.195&2.968&3.839&~~~ &~~\\
~~~& C.K.($n_{f}$=5) \cite{pot5} &0.877&1.718&2.561&3.429&4.406&~~~~~&~~~~~\\
~~~& C.K.($n_{f}$=6) \cite{pot5} &1.105&2.099&3.071&4.057&5.158&~~~~~&~~~~~\\
~~~~& Cor.~~~~~~~~~~ \cite{pot1}&0.732&1.632&2.643&3.753&5.164&~~~&~~ \\
\hline
\end{tabular}
\label{tabrpc}
\end{table}
\end{center}
\end{widetext}

The other input parameters are chosen as the following values \cite{wtd,pdg}: $m_{W}$=80.399 GeV, $m_t = 172.0$ GeV, $|V_{tb}|$=0.88. Leading-order $\alpha_s$ running is adopted and we set the renormalization scale to be $m_{Bc}$ for $|(b\bar{c})\rangle$ quarkonium, which leads to $\alpha_s(m_{Bc})$=0.26, and $2m_b$ for $|(b\bar{b})\rangle$ quarkonium, which leads to $\alpha_s(2m_b)$=0.18. Furthermore, similarly to our previous treatment \cite{wbc2}, we adopt the same constituent quark mass for the same $n$ th-level Fock states \cite{pdg,pot5,sr,tse}. To ensure the gauge invariance of the hard amplitude, we set the $|(b\bar{Q})[n]\rangle$ quarkonium mass $M$ to be $m_b+m_{Q}$.

\subsection{Heavy quarkonium production via top decays}

As a reference, we calculate the decay width for the basic processes $t\to b+W^{+}$. Their decay width can be written as
\begin{eqnarray}
\Gamma &=&\frac {|V_{tb}|^2 G_F |p|} {2 \sqrt{2} \pi {m_t}}(3 {m^2_W} \sqrt{{m_b}^2 +|p|^2}+\nonumber\\
 &&2 |p|^2 (\sqrt {{m^2_b}+|p|^2}+\sqrt{{m^2_W}+|p|^2})).
\end{eqnarray}
where $p$ stands for the relative momentum between the final two particles in the rest frame of the top quark:
\begin{displaymath}
|p| = \frac{\sqrt{(m_t^2 -(m_b -m_W)^2)(m_t^2 -(m_b + m_W)^2)}}{2 m_t}.
\end{displaymath}
Then, we obtain $\Gamma_{t\to b+W^{+}} = 1.131$GeV.

The decay widths for the aforementioned quarkonium states through the production channel, $t\rightarrow |(b\bar{Q})[n]\rangle+QW^{+}$, are listed in Table \ref{tabrpd} with the B.T. potential. Moreover, it must be pointed out that our numerical results for the color-singlet $1S$, $1^1P_1$, and $1^3P_J$ wave $(J=0, 1, 2)$ cases agree with those of Ref.~\cite{tbc2} under the same input values for $t\to |(b\bar{c})[n]\rangle+cW^{+}$.
\begin{widetext}
\begin{center}
\begin{table}
\caption{Decay widths (in keV) for the production of $|(b\bar{Q})[nS]\rangle$ and $|(b\bar{Q})[nP]\rangle$ quarkonium through top decays under the B.T. potential \cite{pot2,wgs,ec}.}
\begin{tabular}{|c|c|c|c|c|c|c|c|c|}
\hline\hline
~~~~~~~~~~~~~~~~~~~ &~~~$n=1$~~~&~~~$n=2$~~~&~~~$n=3$~~~&~~~$n=4$~~~&~~~$n=5$~~~&~~~$n=6$~~~&~~~$n=7$~~~\\
\hline
$\Gamma(t\rightarrow |(b\bar{c})[n^1S_0]\rangle +cW^{+})~~(n_{f}=3)$ &1055& 270.8&148.2&111.8&90.68&~~~&~~~\\
\hline
$\Gamma(t\rightarrow |(b\bar{c})[n^3S_1]\rangle +cW^{+})~~(n_{f}=3)$ &1473& 356.3&192.2&142.8&115.9&~~~&~~~\\
\hline
$\Gamma(t\rightarrow |(b\bar{c})[n^1P_1]\rangle +cW^{+})~~(n_{f}=3)$ &48.33&26.45&24.35&21.37&~~~&~~~&~~~\\
\hline
$\Gamma(t\rightarrow |(b\bar{c})[n^3P_0]\rangle +cW^{+})~~(n_{f}=3)$ &88.81&54.36&55.38&50.79&~~~&~~~&~~~\\
\hline
$\Gamma(t\rightarrow |(b\bar{c})[n^3P_1]\rangle +cW^{+})~~(n_{f}=3)$ &30.19&16.97&16.02&14.23&~~~&~~~&~~~\\
\hline
$\Gamma(t\rightarrow |(b\bar{c})[n^3P_2]\rangle +cW^{+})~~(n_{f}=3)$ &28.32&14.97&26.99&11.53&~~~&~~~&~~~\\
\hline\hline
$\Gamma(t\rightarrow |(b\bar{b})[n^1S_0]\rangle +bW^{+})~~(n_{f}=4)$ &57.63&18.43&5.345&5.974&5.635&5.152&4.529\\
\hline
$\Gamma(t\rightarrow |(b\bar{b})[n^3S_1]\rangle +bW^{+})~~(n_{f}=4)$ &56.85&18.10&5.238&5.847&5.504&5.026&4.413\\
\hline
$\Gamma(t\rightarrow |(b\bar{b})[n^1P_1]\rangle +cW^{+})~~(n_{f}=3)$ &1.961&0.776&0.650&0.679&0.681&0.639&~~~\\
\hline
$\Gamma(t\rightarrow |(b\bar{b})[n^3P_0]\rangle +cW^{+})~~(n_{f}=3)$ &13.05&5.24&4.420&4.674&4.743&4.482&~~~\\
\hline
$\Gamma(t\rightarrow |(b\bar{b})[n^3P_1]\rangle +cW^{+})~~(n_{f}=3)$ &1.670&0.664&0.561&0.589&0.595&0.561&~~~\\
\hline
$\Gamma(t\rightarrow |(b\bar{b})[n^3P_2]\rangle +cW^{+})~~(n_{f}=3)$ &0.547&0.219&0.183&0.191&0.193&0.181&~~~\\
\hline
\end{tabular}
\label{tabrpd}
\end{table}
\end{center}
\end{widetext}

From Table \ref{tabrpd}, it is found that, in addition to the ground $1S$-level states, the higher $|(b\bar{Q})[n]\rangle$  quarkonium states can also provide sizable contributions to the total decay widths. For convenience, we have used $[nS]$ to present the summed decay widths of $[n^1S_0]$ and $[n^3S_1]$ at the same $n$th level, and $[nP]$ to represent the summed decay widths of $[n^1P_1]$ and $[n^3P_J] (J=0, 1, 2)$ at the same $n$th level.
\begin{itemize}
\item For $|(b\bar{c})[n]\rangle$ quarkonium production through the channel $t\rightarrow |(b\bar{c})[n]\rangle + cW^{+}$, the total decay widths for all $2S$, $3S$, $4S$, $5S$, $1P$, $2P$, $3P$, and $4P$-wave states is $24.8\%$, $13.5\%$, $10.1\%$, $8.2\%$, $7.7\%$, $4.5\%$, $4.8\%$, and $3.9\%$ of those of $B_c$ and $B^*_c$. Considering that the LHC runs at the center-of-mass energy $\sqrt{S}=14$ TeV with the luminosity ${\cal L}\propto 10^{34}cm^{-2}s^{-1}$, one expects that about $1.0\times10^8$ events per year can be generated. Then we can estimate the heavy quarkonium events generated through top quark decays; i.e., $2.2\times10^5$  $|(b\bar{c})[1S]\rangle$, $5.5\times10^4$ $|(b\bar{c})[2S]\rangle$, $3.0\times10^4$ $|(b\bar{c})[3S]\rangle$, $2.3\times10^4$ $|(b\bar{c})[4S]\rangle$, $1.8\times10^4$ $|(b\bar{c})[5S]\rangle$, $1.7\times10^4$ $|(b\bar{c})[1P]\rangle$, $1.0\times10^4$ $|(b\bar{c})[2P]\rangle$, $1.1\times10^4$ $|(b\bar{c})[3P]\rangle$, and $8.7\times10^3$ $|(b\bar{c})[4P]\rangle$ quarkonium events per year can be obtained.
\end{itemize}

\begin{itemize}
\item For $|(b\bar{b})[n]\rangle$ quarkonium production, the total decay widths for all $2S$, $3S$, $4S$, $5S$, $6S$, $7S$, $1P$, $2P$, $3P$, $4P$, $5P$, and $6P$ wave states are about $31.9\%$, $9.2\%$, $10.3\%$ $9.7\%$, $8.9\%$, $7.8\%$, $15.0\%$, $6.0\%$, $5.1\%$, $5.4\%$, $5.4\%$, and $5.1\%$ of those of $\eta_b$ and $\Upsilon$ for $t\rightarrow |(b\bar{b})[n]\rangle+ bW^{+}$. At the LHC, i.e.,  $1.0\times10^4$  $|(b\bar{b})[1S]\rangle$, $3.2\times10^3$ $|(b\bar{b})[2S]\rangle$, $9.4\times10^2$ $|(b\bar{b})[3S]\rangle$, $1.0\times10^3$ $|(b\bar{b})[4S]\rangle$, $9.8\times10^2$ $|(b\bar{b})[5S]\rangle$, $9.0\times10^2$ $|(b\bar{b})[6S]\rangle$, $7.9\times10^2$ $|(b\bar{b})[7S]\rangle$, and summed up, $4.3\times10^3$ $|(b\bar{b})[P]\rangle$ quarkonium events per year can be obtained.
\end{itemize}

To show the relative importance among  different Fock states more clearly, we present the differential distributions
$d\Gamma/ds_{1}$, $d\Gamma/ds_{2}$,  $d\Gamma/dcos\theta_{12}$, and $d\Gamma/dcos\theta_{13}$ for the mentioned channels in Figs. \ref{tW(bc)cds1ds2sum}, \ref{tW(bb)bds1ds2sum}, \ref{tW(bc)cdscos12sum} and \ref{tW(bb)bdcos13sum}. Moreover, we define a ratio
\begin{eqnarray}
R_i[n]=\frac {d\Gamma/ds_i(|(b\bar{Q})[n]\rangle)} {d\Gamma/ds_i(|(b\bar{Q})[1S]\rangle)},
\end{eqnarray}
where $i=1, 2$ and $n=2S, 3S, 1P$, and $2P$. The curves are presented in Fig.~\ref{RtBcWds12}. These figures show explicitly that the higher Fock states $|(b\bar{c})[2S]\rangle$, $|(b\bar{c})[3S]\rangle$, $|(b\bar{c})[1P]\rangle$, and $|(b\bar{c})[2P]\rangle$ can provide sizable contributions in comparison to the lower Fock state $|(b\bar{c})[1S]\rangle$ in almost the entire kinematical region.

If all of the higher excited heavy quarkonium states decay to the ground spin-singlet $S$ wave state $|(b\bar{Q})[1^1S_0]\rangle$ with $100\%$ efficiency via electromagnetic or hadronic interactions, then we obtain the total decay width of top quark decay channels within the B.T. potential model:
\begin{eqnarray}
\Gamma{(t\to |(b\bar{c})[1^1S_0]\rangle +cW^{+})} &=&4486\;{\rm KeV} \label{bct} ,\\
\Gamma{(t\to |(b\bar{b})[1^1S_0]\rangle +bW^{+})} &=&251.8\;{\rm KeV} \label{bbt1}.
\end{eqnarray}

At the LHC, running at the center-of-mass energy $\sqrt{S}=14$ TeV with luminosity $10^{34} cm^{-2} s^{-1}$, one may expect to produce about $10^8$ $t\bar{t}$-pairs per year \cite{bc1,bc2}. Then we can estimate the event number of $|(b\bar{Q})\rangle$ quarkonium production through top quark decays, i.e., $2.0\times10^5$ $|(b\bar{c})\rangle$ quarkonium events and $1.0\times10^4$ $|(b\bar{b})\rangle$ quarkonium events per year. It might be possible to find $B_c$ and $\Upsilon$ through top quark decays, since one may identify these particles through their cascade decay channels, $B_c \to J/\psi+\pi$ or $B_c\to J/\psi +e\nu_e$ with clear signals. Bearing in mind the situation pointed out here and the possible upgrade for the LHC (SLHC, DLHC, etc. \cite{ab}), the possibility to study $|(b\bar{c})\rangle$ quarkonium and bottomonium via top quark decays is worth thinking seriously about.

\begin{widetext}
\begin{center}
\begin{figure}
\includegraphics[width=0.45\textwidth]{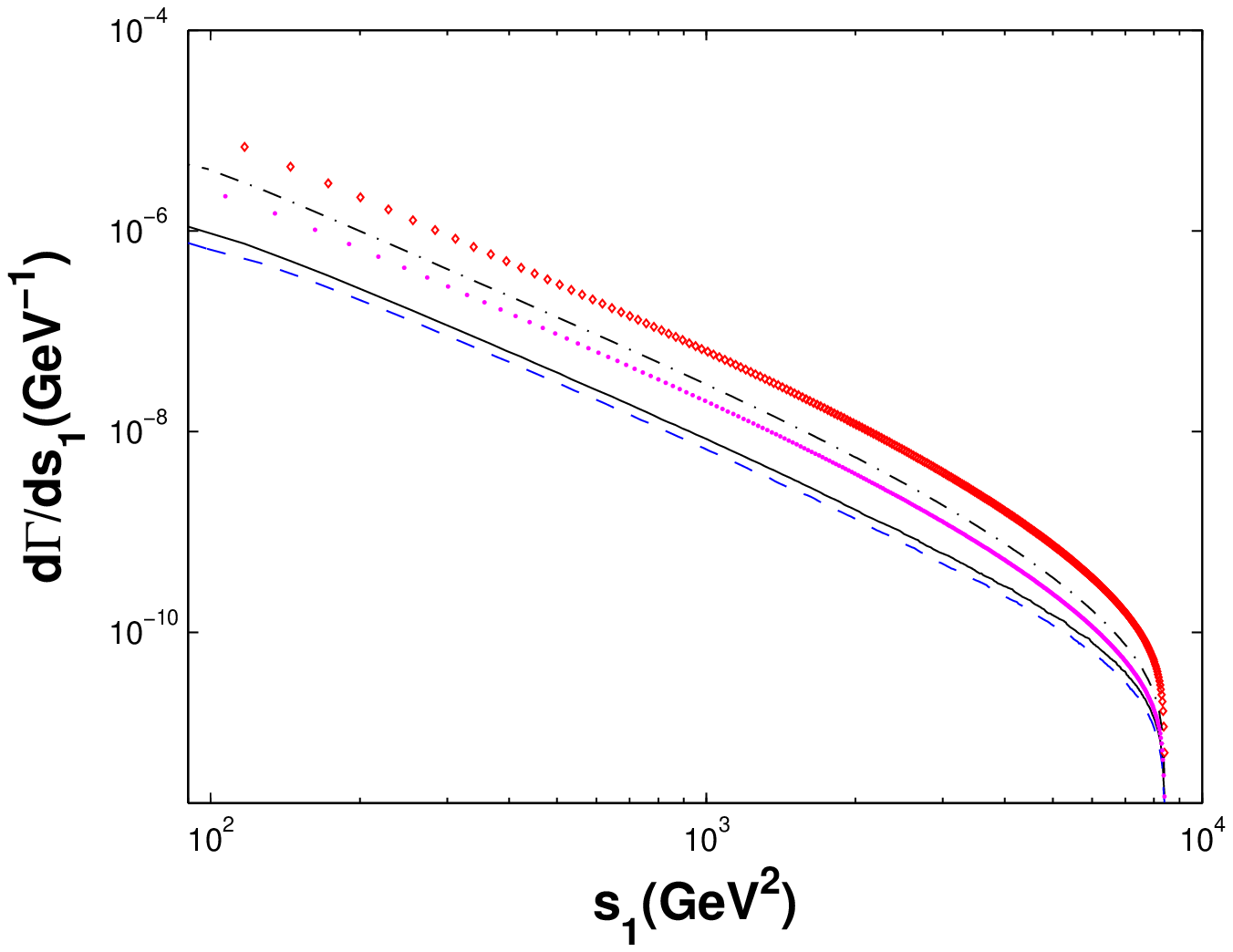}
\includegraphics[width=0.45\textwidth]{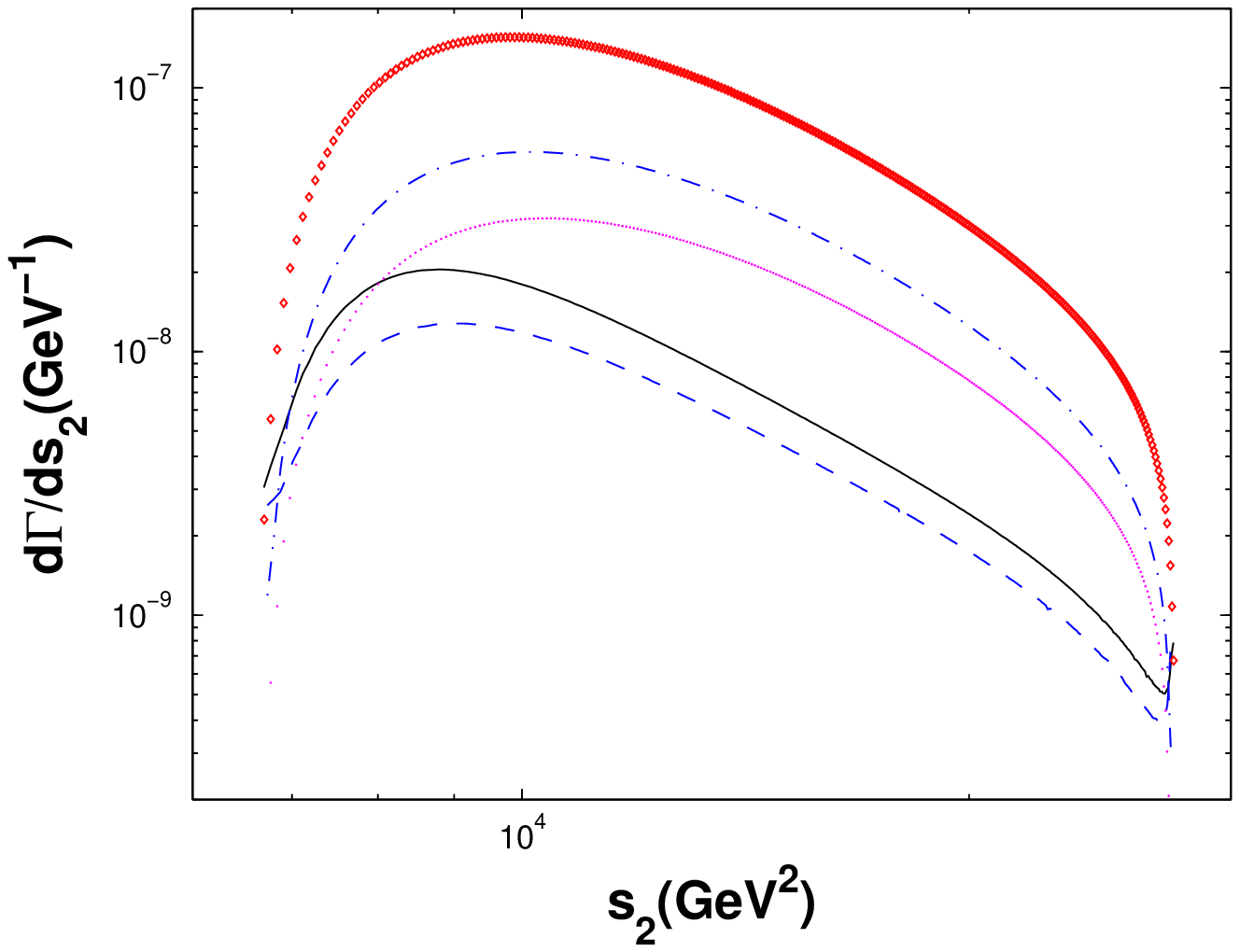}
\caption{Differential decay widths $d\Gamma/ds_1$ and $d\Gamma/ds_2$ for $ t\rightarrow |(b\bar{c})[n]\rangle +cW^{+}~(n_{f}=3)$, where the diamond line, the dash-dotted line, the dotted line, the solid line, and the dashed line are for $|(b\bar{c})[1S]\rangle$,  $|(b\bar{c})[2S]\rangle$, $|(b\bar{c})[3S]\rangle$, $|(b\bar{c})[1P]\rangle$, and $|(b\bar{c})[2P]\rangle$, respectively.} \label{tW(bc)cds1ds2sum}
\end{figure}
\end{center}

\begin{center}
\begin{figure}
\includegraphics[width=0.45\textwidth]{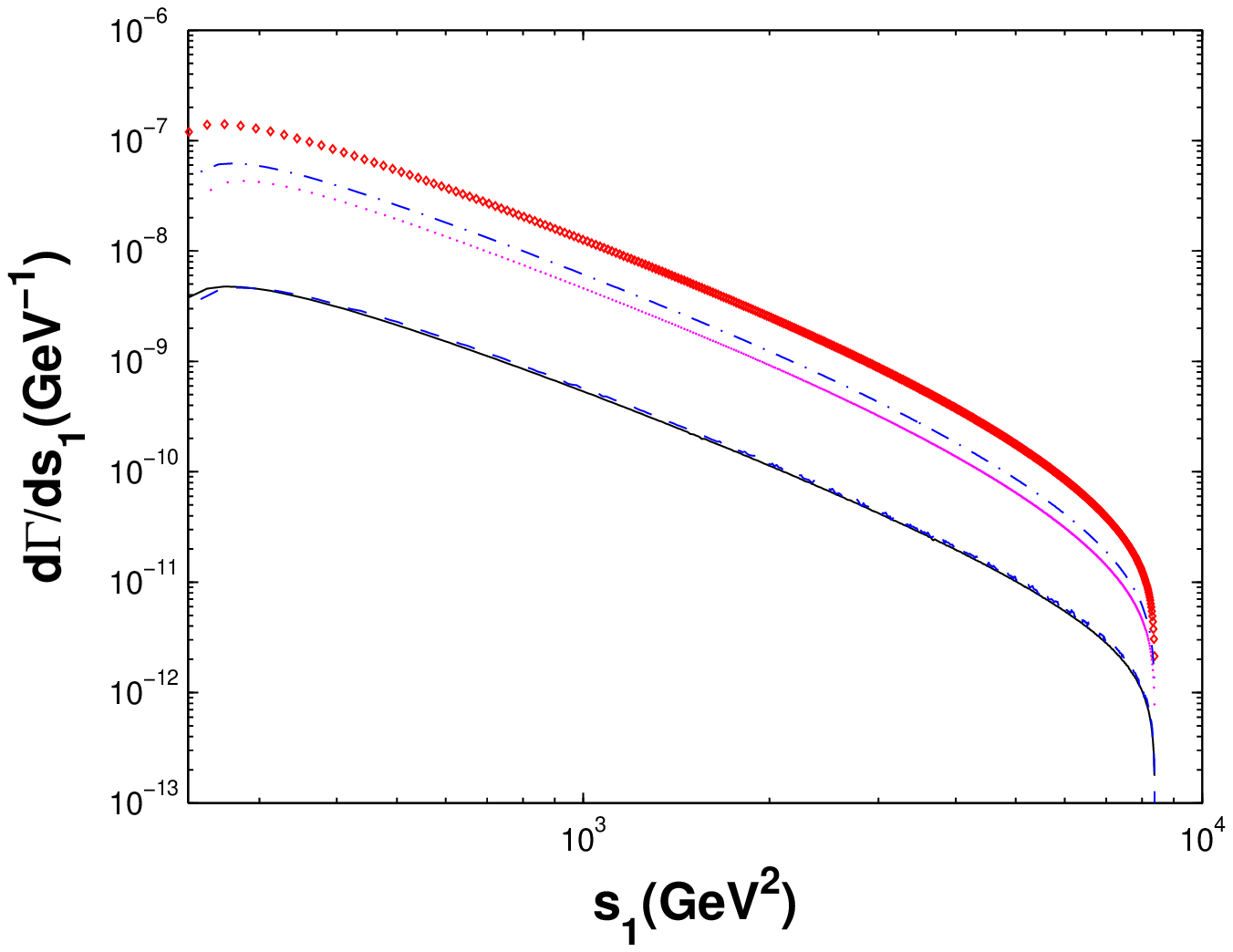}
\includegraphics[width=0.45\textwidth]{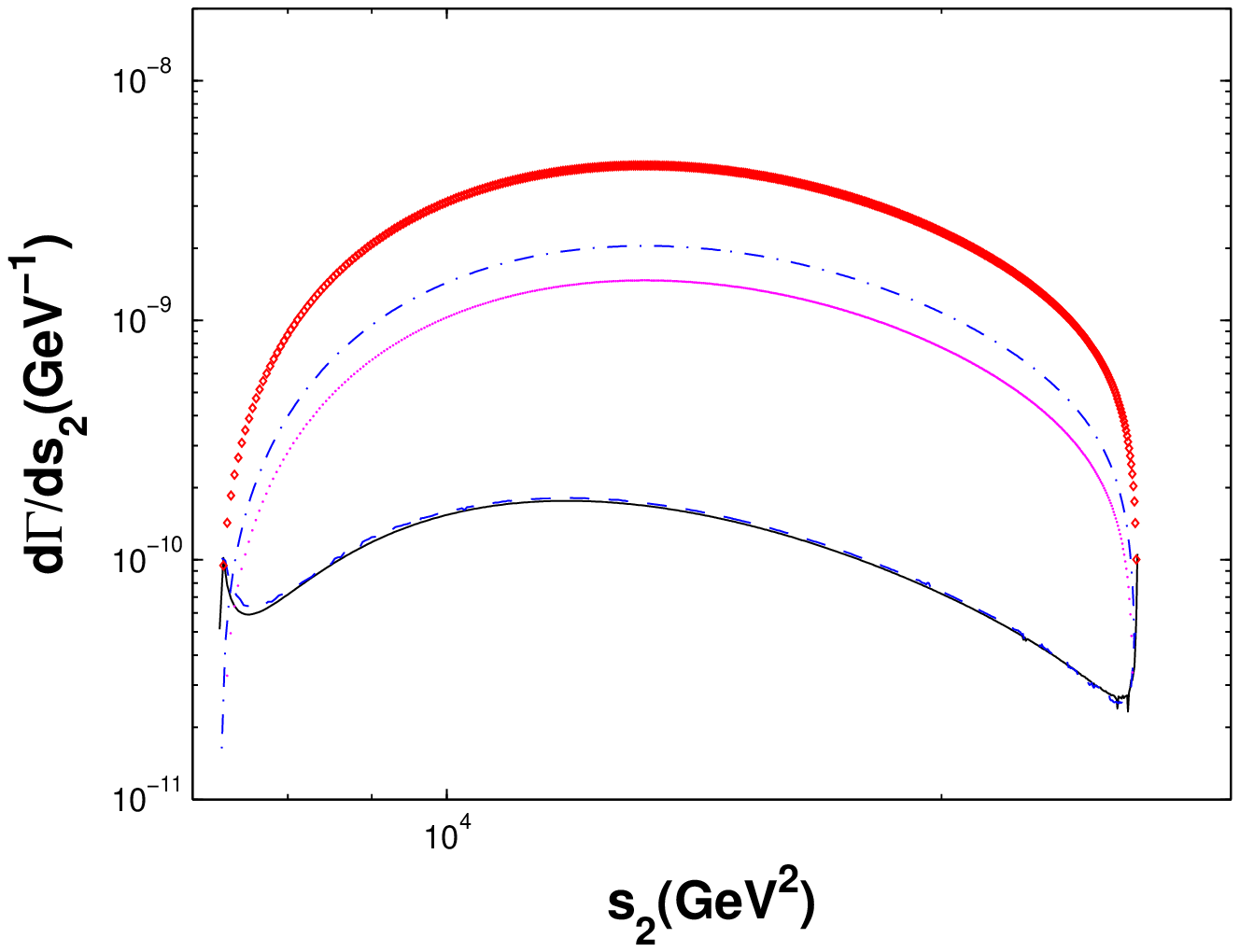}
\caption{Differential decay widths $d\Gamma/ds_1$ and $d\Gamma/ds_2$ for $ t\rightarrow |(b\bar{b})[n]\rangle +bW^{+}~(n_{f}=4)$, where the diamond line, the dash-dotted line, the dotted line, the solid line, and the dashed line are for $|(b\bar{b})[1S]\rangle$,  $|(b\bar{b})[2S]\rangle$, $|(b\bar{b})[3S]\rangle$, $|(b\bar{b})[1P]\rangle$, and $|(b\bar{b})[2P]\rangle$, respectively.} \label{tW(bb)bds1ds2sum}
\end{figure}
\end{center}

\begin{center}
\begin{figure}
\includegraphics[width=0.45\textwidth]{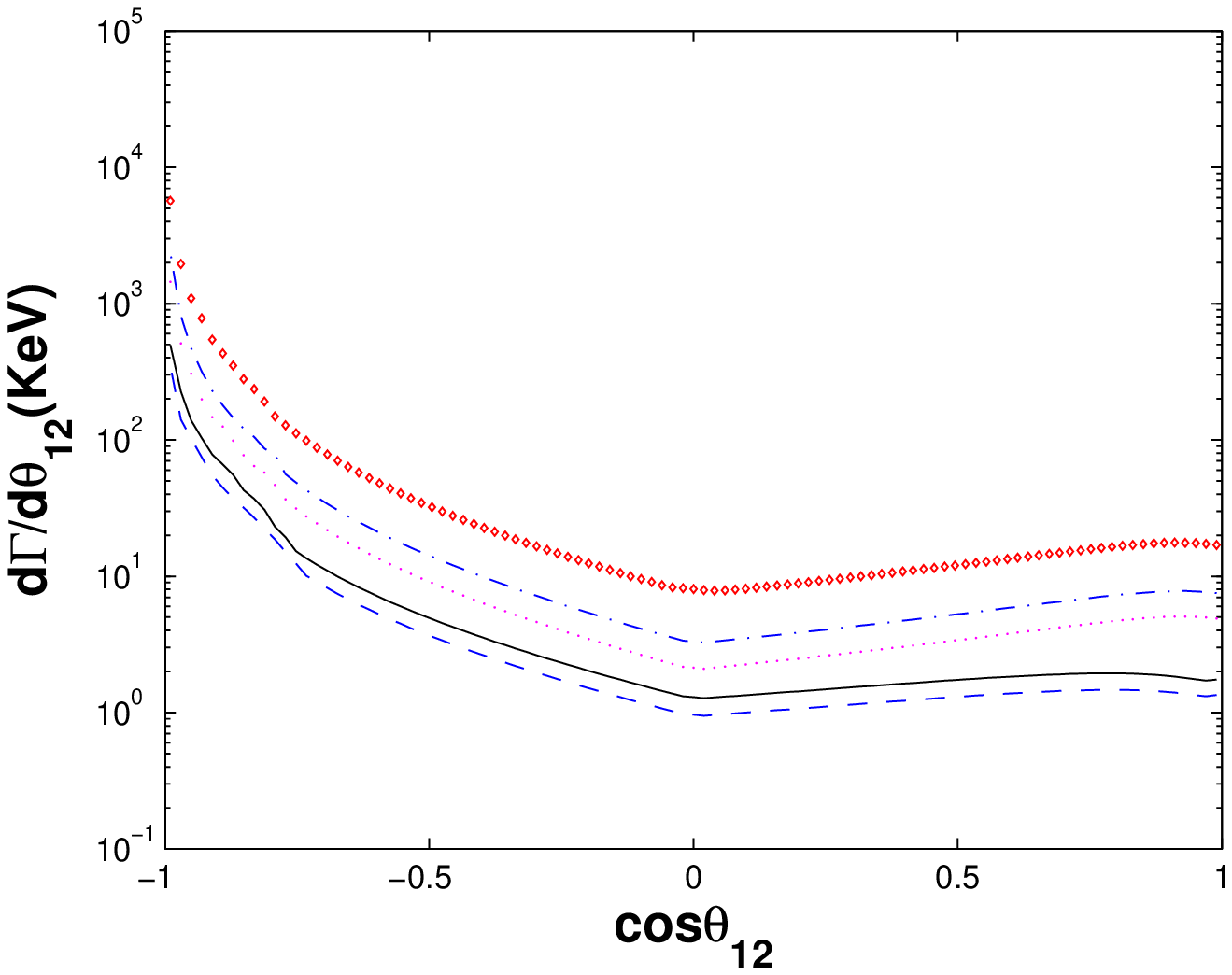}
\includegraphics[width=0.45\textwidth]{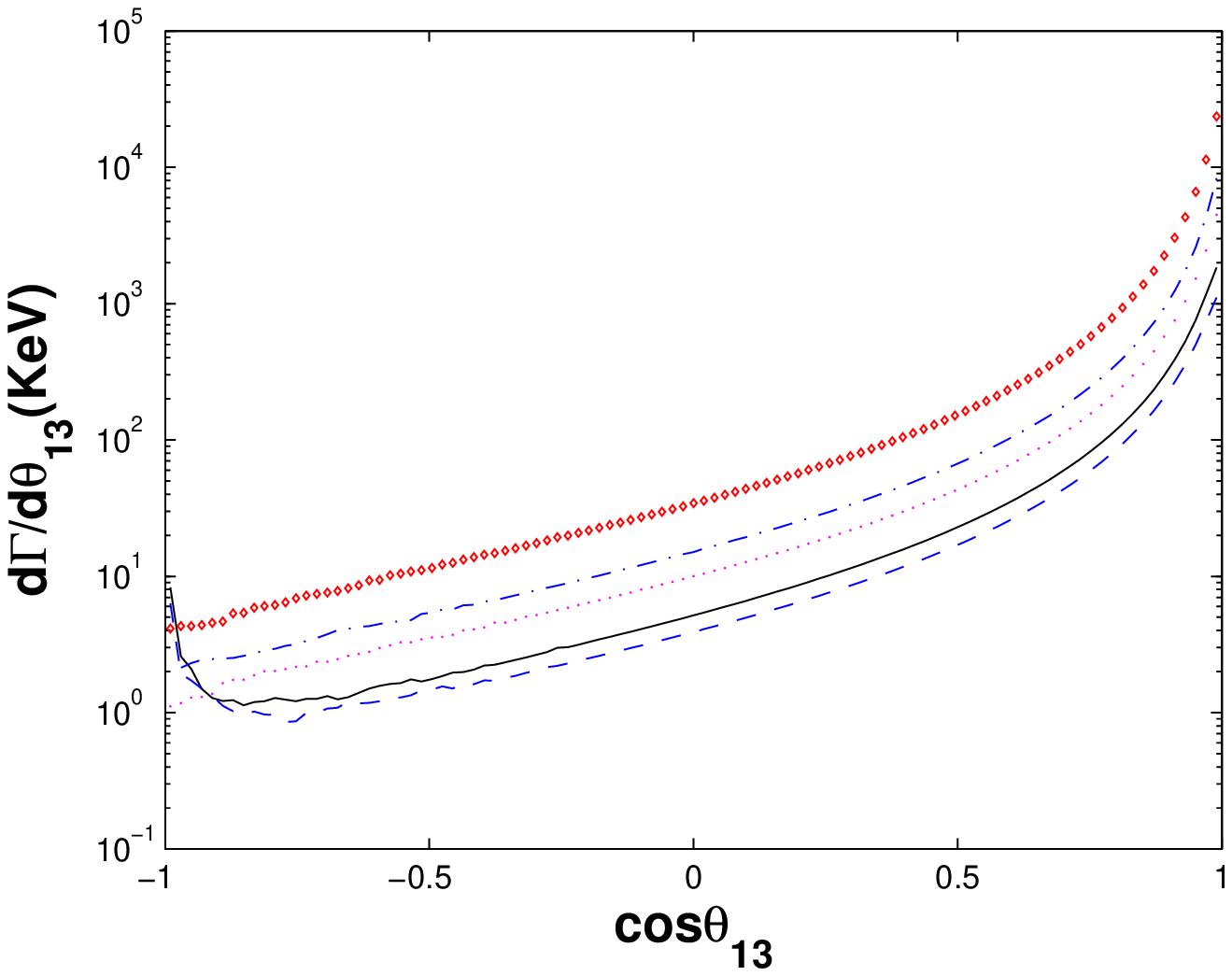}
\caption{Differential decay widths $d\Gamma/dcos\theta_{12}$ and $d\Gamma/dcos\theta_{13}$ for $ t\rightarrow |(b\bar{c})[n]\rangle +cW^{+}~(n_{f}=3)$, where the diamond line, the dash-dotted line, the dotted line, the solid line, and the dashed line are for $|(b\bar{c})[1S]\rangle$,  $|(b\bar{c})[2S]\rangle$, $|(b\bar{c})[3S]\rangle$, $|(b\bar{c})[1P]\rangle$, and $|(b\bar{c})[2P]\rangle$, respectively.} \label{tW(bc)cdscos12sum}
\end{figure}
\end{center}

\begin{center}
\begin{figure}
\includegraphics[width=0.45\textwidth]{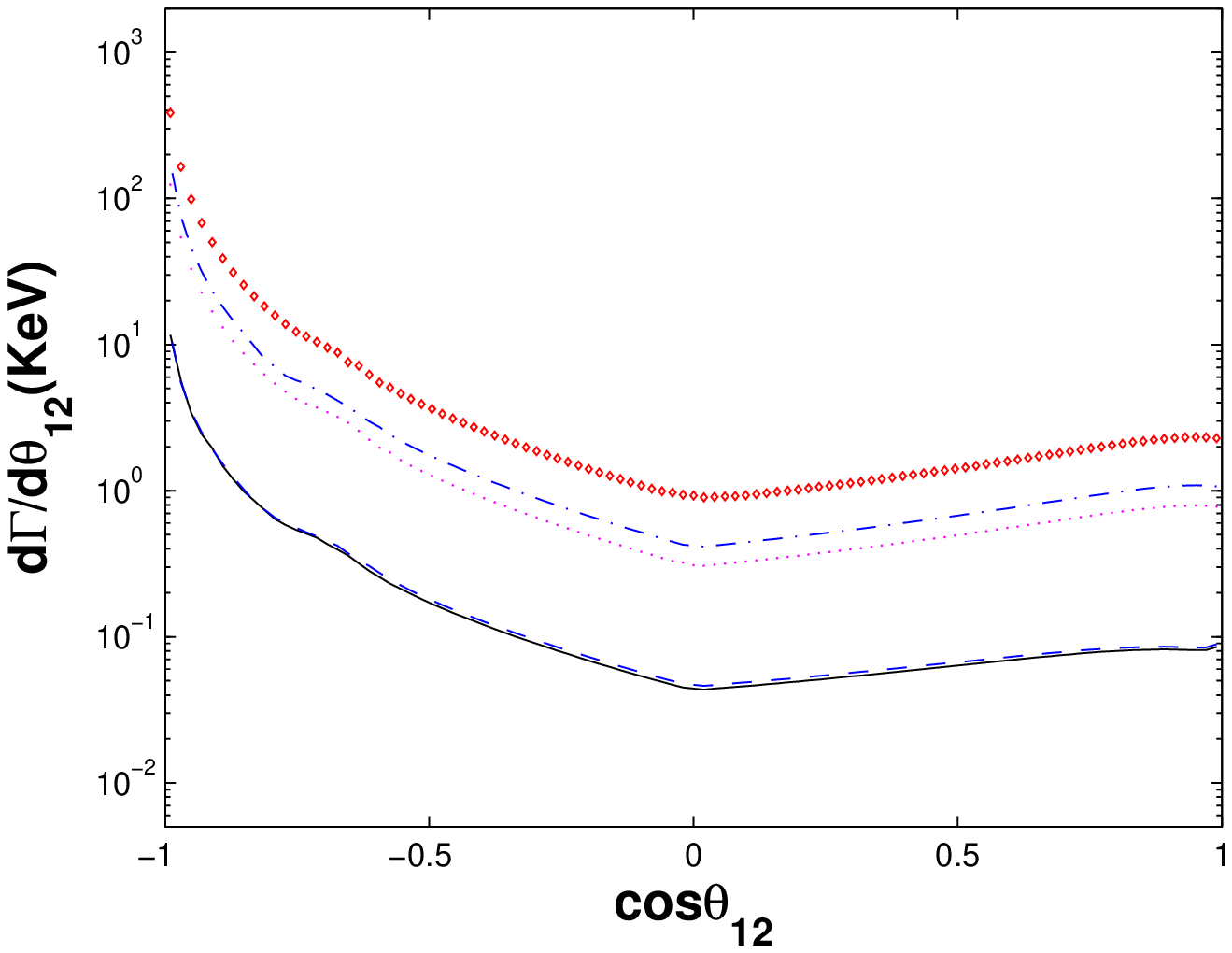}
\includegraphics[width=0.45\textwidth]{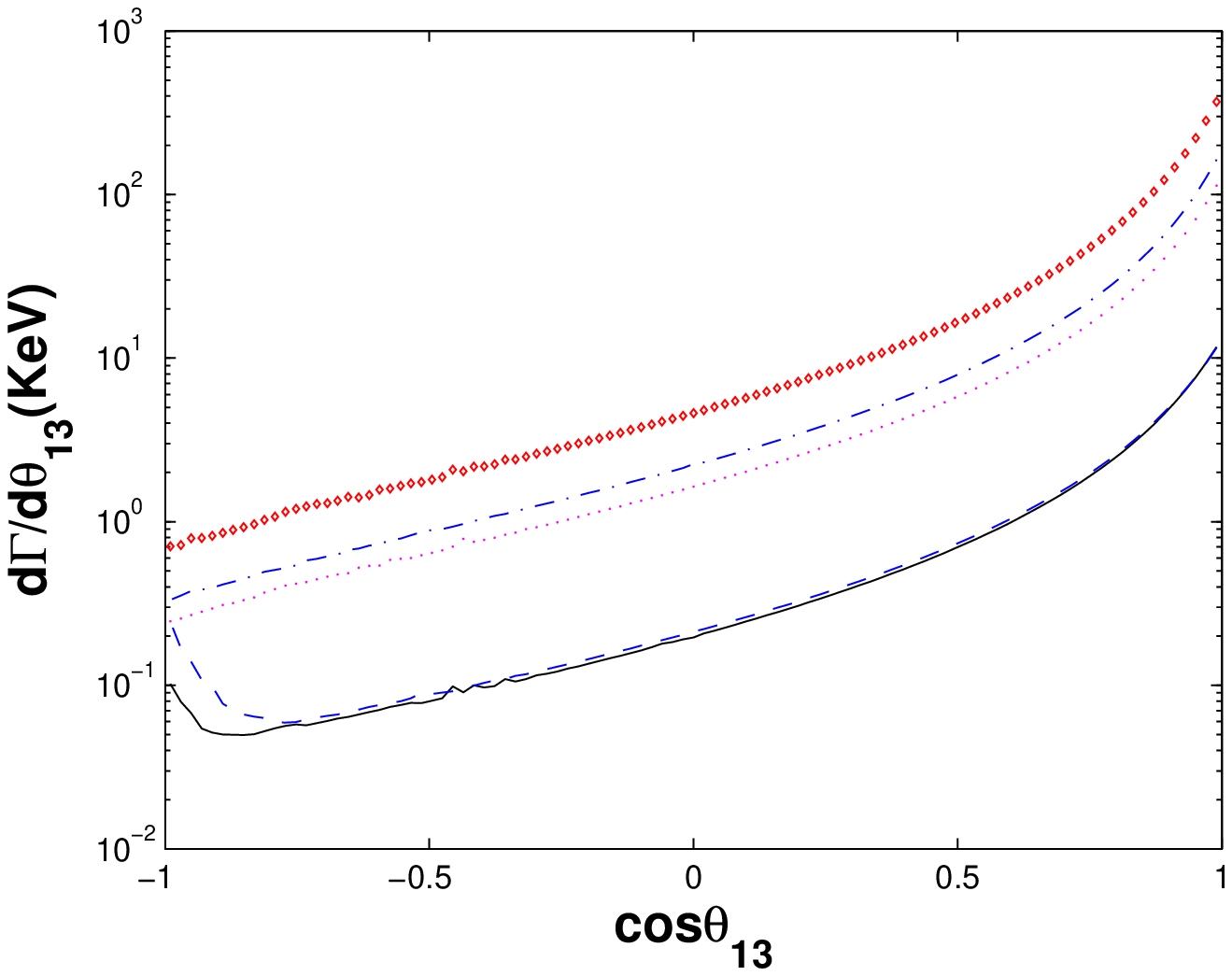}
\caption{Differential decay widths $d\Gamma/dcos\theta_{12}$ and $d\Gamma/dcos\theta_{13}$ for $ t\rightarrow |(b\bar{b})[n]\rangle +bW^{+}~(n_{f}=4)$, where the diamond line, the dash-dotted line, the dotted line, the solid line, and the dashed line are for $|(b\bar{b})[1S]\rangle$,  $|(b\bar{b})[2S]\rangle$, $|(b\bar{b})[3S]\rangle$, $|(b\bar{b})[1P]\rangle$, and $|(b\bar{b})[2P]\rangle$, respectively.} \label{tW(bb)bdcos13sum}
\end{figure}
\end{center}

\begin{center}
\begin{figure}
\includegraphics[width=0.45\textwidth]{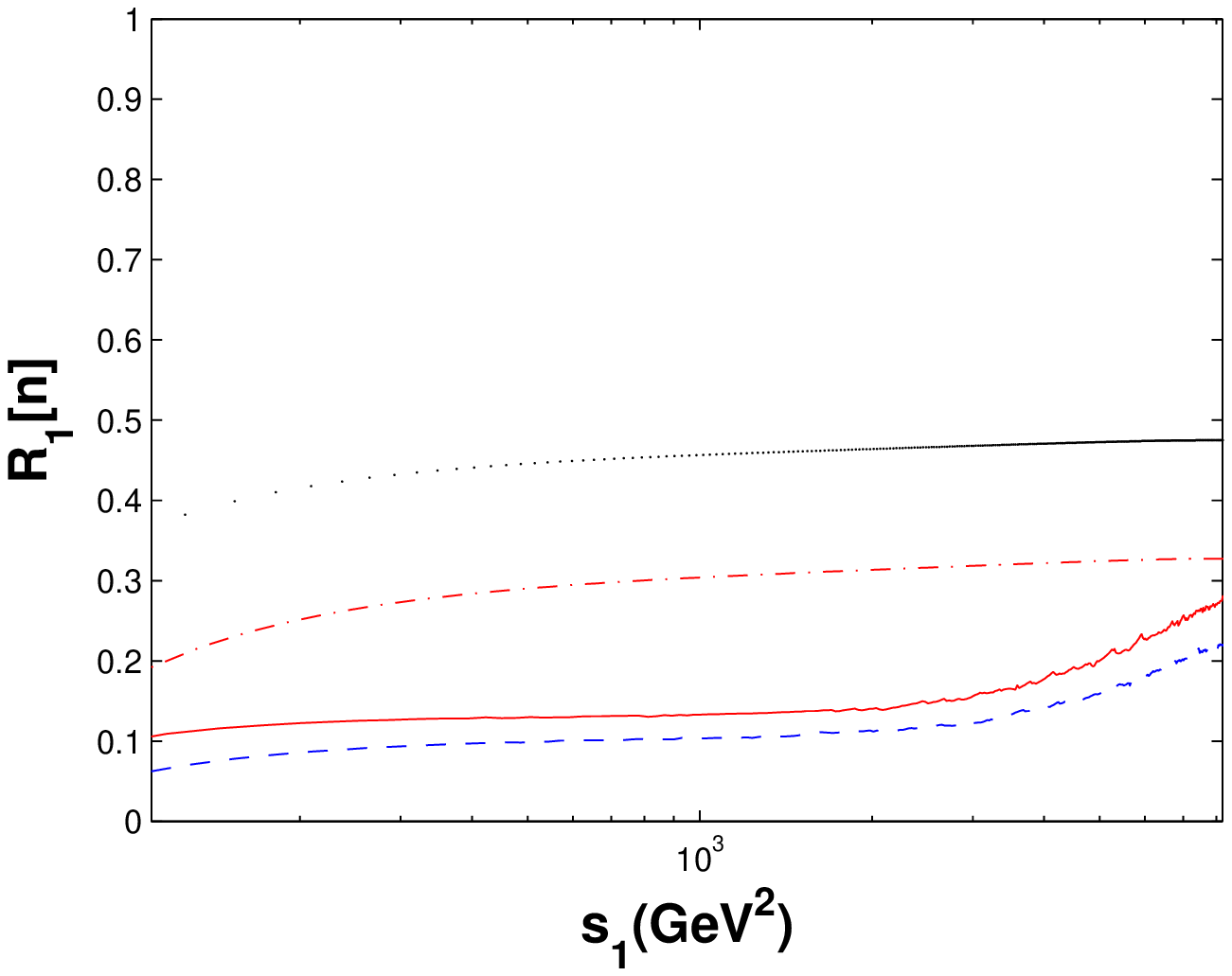}
\includegraphics[width=0.45\textwidth]{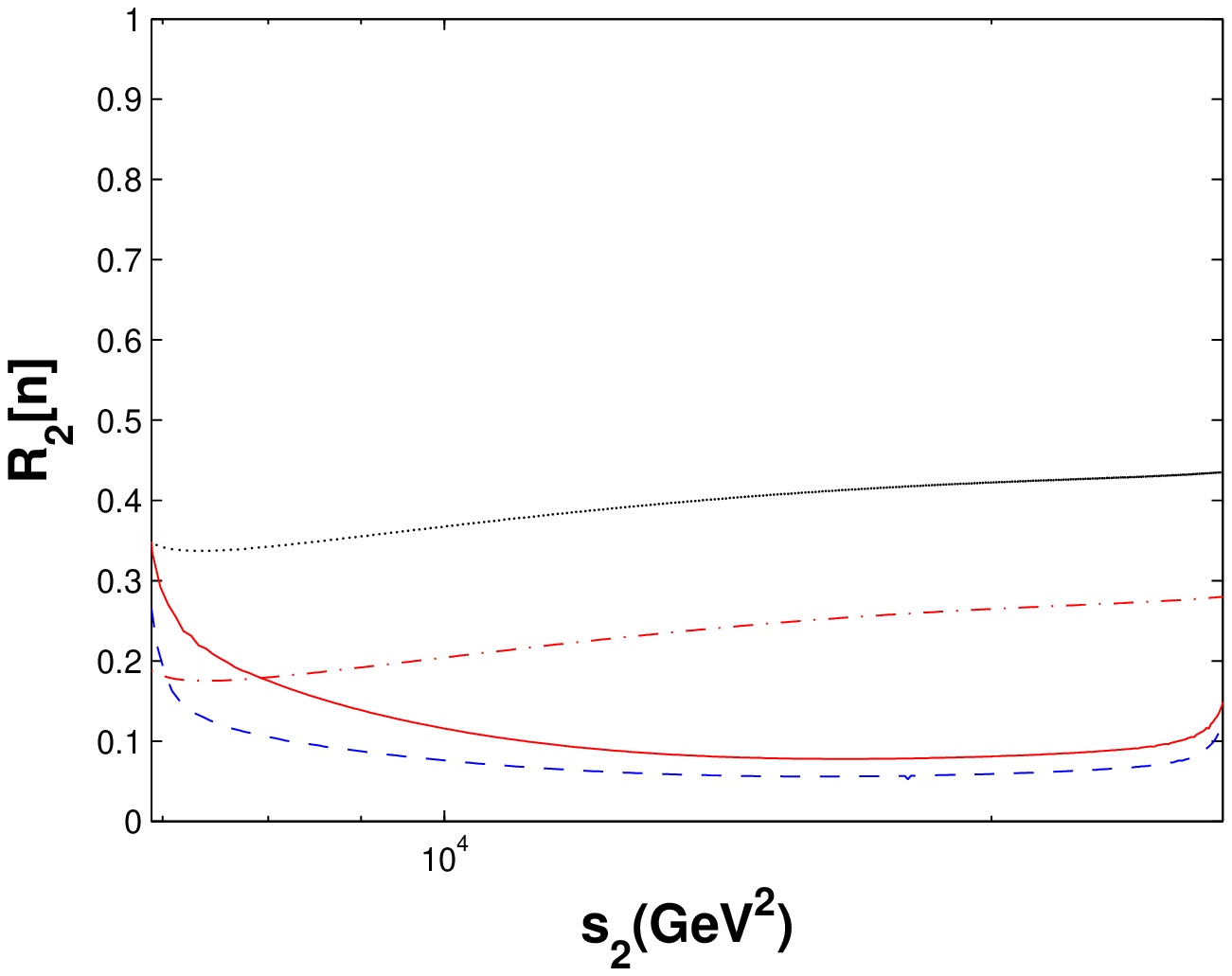}
\caption{The ratios $R_1[n]$ and $R_2[2]$ versus $s_1$ and $s_2$ for the channel $t\rightarrow |(b\bar{c})[n]\rangle+cW^{+}~(n_{f}=3)$. Here the dotted line, the dash-dotted line, the solid line, and the dashed line are for $|(b\bar{c})[2S]\rangle$, $|(b\bar{c})[3S]\rangle$, $|(b\bar{c})[1P]\rangle$, and $|(b\bar{c})[2P]\rangle$, respectively.} \label{RtBcWds12}
\end{figure}
\end{center}
\end{widetext}

\subsection{Decay widths under five potential models}

In this subsection, we discuss the uncertainties caused by the bound-state parameters. These parameters are the main uncertainty source for estimating heavy $|(b\bar{Q})[n]\rangle$ quarkonium production. In this paper, we discuss the decay widths of $|(b\bar{c})[n]\rangle$ quarkonium and bottomonium production through top quark decays under five potential models in detail, i.e., the B.T. potential \cite{pot2}, the J. potential \cite{jlr}, the I.O. potential \cite{kso,sr}, the C.K. potential \cite{pot5,sr}, and the Cor. model \cite{pot1}. The constituent quark masses and their corresponding radial wave functions at the origin and the first derivative of the radial wave function at the origin for the $|(b\bar{c})[n]\rangle$ ($n_{f}$=3) and $|(b\bar{b})[n]\rangle$ quarkonium ($n_{f}$=4)  states can be adopted in Tables~\ref{tabrpb} and~\ref{tabrpc}.

The decay widths for $|(b\bar{c})[n]\rangle$  and $|(b\bar{b})[n]\rangle$ quarkonium production under five potential models are presented in Tables~\ref{tabrpe} and~\ref{tabrpf}. The decay widths for the five models are consistent with each other: taking the B.T. model decay width as the center value,  for the channel $t\rightarrow |(b\bar{c})[n]\rangle + cW^{+}$, we obtain the uncertainty $(^{+46\%}_{-47\%})$, where the upper value is from the O.I. model and the lower value is from the C.H. model; and the uncertainty $(^{+0\%}_{-56\%})$ for the channel $t\rightarrow |(b\bar{b})[n]\rangle + bW^{+}$, where the lower value is from the C.K. model.

\begin{table}
\caption{Decay widths (in keV) for $|(b\bar{c})[n]\rangle$ quarkonium production channel $t\rightarrow |(b\bar{c})[n]\rangle+cW^{+}~(n_{f}=3)$, where bound-state parameters from five potential models are adopted.}
\begin{tabular}{|c||c|c|c|c|c|c|c|}
\hline
~~~~&B.T. \cite{pot2}&J. \cite{jlr}&I.O. \cite{kso}&C.K. \cite{pot5}&Cor. \cite{pot1}\\
\hline\hline
$[n]=[1^1S_0]$ &1055&554.1&1703&357.5&488.8\\
\hline
$[n]=[1^3S_1]$ &1473&773.6&2378&499.2&682.5\\
\hline
$[n]=[2^1S_0]$ &270.8&495.3&595.2&287.1&437.5\\
\hline
$[n]=[2^3S_1]$ &356.3&651.6&783.2&377.7&575.6\\
\hline
$[n]=[3^1S_0]$ &148.2&182.3&143.2&102.7&161.2\\
\hline
$[n]=[3^3S_1]$ &192.2&236.3&185.7&133.2&209.0\\
\hline
$[n]=[4^1S_0]$ &111.8&141.9&82.21&78.91&126.0\\
\hline
$[n]=[4^3S_1]$ &142.8&181.3&105.0&100.8&161.0\\
\hline
$[n]=[5^1S_0]$ &90.68&127.4&59.53&70.40&113.0\\
\hline
$[n]=[5^3S_1]$ &115.9&162.8&76.09&89.99&144.4\\
\hline
$[n]=[1P]$ &195.7&156.0&216.5&70.27&82.74\\
\hline
$[n]=[2P]$&112.8&147.8&104.2&67.32&81.93\\
\hline
$[n]=[3P]$ &122.7&158.8&69.98&65.61&90.45\\
\hline
$[n]=[4P]$ &97.96&137.3&43.31&59.36&79.45\\
\hline
Sum. &4486&4107&6545&2360&3434\\
\hline
\end{tabular}
\label{tabrpe}
\end{table}

\begin{table}
\caption{Decay widths (in keV) for the $|(b\bar{b})[n]\rangle$ quarkonium production channel $t\rightarrow |(b\bar{b})[n]\rangle+bW^{+}$~($n_{f}$=4), where bound-state parameters from five potential models are adopted.}
\begin{tabular}{|c||c|c|c|c|c|c|c|}
\hline
~~~~~&B.T. \cite{pot2}&J. \cite{jlr}&I.O. \cite{kso}&C.K. \cite{pot5}&Cor. \cite{pot1}\\
\hline\hline
$[n]=[1^1S_0]$ &57.63&23.83&33.44&17.75&30.62\\
\hline
$[n]=[1^3S_1]$ &56.85&23.50&32.98&17.50&30.20\\
\hline
$[n]=[2^1S_0]$ &18.43&11.33&9.455&7.604&13.03\\
\hline
$[n]=[2^3S_1]$ &18.10&11.13&9.293&7.469&12.80\\
\hline
$[n]=[3^1S_0]$ &5.345&8.371&5.047&5.464&9.603\\
\hline
$[n]=[3^3S_1]$ &5.238&8.201&4.947&5.355&9.409\\
\hline
$[n]=[4^1S_0]$ &5.974&7.034&3.358&5.974&8.077\\
\hline
$[n]=[4^3S_1]$ &5.847&6.884&3.286&4.455&7.907\\
\hline
$[n]=[5^1S_0]$ &5.635&6.102&2.417&3.935&7.028\\
\hline
$[n]=[5^3S_1]$ &5.504&5.960&2.361&3.843&6.865\\
\hline
$[n]=[6^1S_0]$ &5.152&5.525&1.883&3.560&6.367\\
\hline
$[n]=[6^3S_1]$ &5.026&5.389&1.836&3.472&6.211\\
\hline
$[n]=[7^1S_0]$ &4.529&5.097&1.528&3.283&5.875\\
\hline
$[n]=[7^3S_1]$ &4.413&4.966&1.488&3.199&5.724\\
\hline
$[n]=[1P]$ &17.23&4.822&3.418&3.259&3.574\\
\hline
$[n]=[2P]$&6.903&5.241&2.357&3.235&4.070\\
\hline
$[n]=[3P]$ &5.814&5.532&1.789&3.270&4.423\\
\hline
$[n]=[4P]$ &6.134&5.417&1.336&3.120&4.433\\
\hline
$[n]=[5P]$ &6.212&5.465&2.473&3.091&4.543\\
\hline
$[n]=[6P]$ &5.863&5.482&1.132&3.060&4.612\\
\hline
Sum. &251.8&165.3&125.8&110.6&185.4\\
\hline
\end{tabular}
\label{tabrpf}
\end{table}

In the present paper, we only calculate and discuss the decay widths of $nS$ and $nP$ wave of $|(b\bar{c})[n]\rangle$ and $|(b\bar{b})[n]\rangle$ quarkonium via the top-quark decays under the five potential models. Yet we believe that the values
of the wave functions at the origin of $nS$, $nP$, and $nD$ wave of $(c\bar{c})$, $(b\bar{c})$, $(b\bar{b})$ in Tables \ref{tabrpa}, \ref{tabrpb}, and \ref{tabrpc} under the five potential models are helpful for both theoretical and experimental study.

\section{Conclusions}

In the present paper, we have calculated the values of the Schr${\rm \ddot{o}}$dinger radial wave function at the origin of $|(c\bar{c})[n]\rangle$, $|(b\bar{c})[n]\rangle$, and $|(b\bar{b})[n]\rangle$ quarkonium for the five potential models, and  made a detailed study on the higher excited heavy quarkonium production through top quark semiexclusive decays, i.e., $t\to |(b\bar{c})[n]\rangle +cW^{+}$ and $t\to |(b\bar{b})[n]\rangle +bW^{+}$, within the NRQCD framework. Results for $|(b\bar{Q})[n]\rangle$ quarkonium Fock states, i.e., $|(b\bar{Q})[n^1S_0]\rangle$ and $|(b\bar{Q})[n^3S_1]\rangle$, and $|(b\bar{Q})[n^1P_1]\rangle$ and $|(b\bar{Q})[n^3P_J]\rangle$ ($n=1 ,\cdots, 6; J=0 ,1 , 2$) have been presented. And to provide the analytical expressions as simply as possible, we have adopted the `improved trace technology' developed in Refs.~\cite{tbc2,zbc0,zbc1,zbc2,wbc1,wbc2} to derive Lorentz- invariant expressions for top quark decay processes at the amplitude level. Such a calculation technology shall be very helpful for dealing with processes with massive spinors.

Numerical results show that higher $nS$ and $nP$ wave states in addition to the  ground $1S$ wave states can also provide sizable contributions to heavy quarkonium production through top quark decays, so one needs to take the higher $nS$ and $nP$ wave states into consideration for a sound estimation. If all the excited states decay to the ground state $|(b\bar{Q})[1^1S_0]\rangle$ with $100\%$ efficiency, we can obtain the total decay width for $|(b\bar{Q})\rangle$ quarkonium production through top quark decays as shown by Eqs.~(\ref{bct}) and (\ref{bbt1}). At the LHC, due to its high collision energy and high luminosity, sizable heavy quarkonium events can be produced through top quark decays, i.e., $2.0\times10^5$ $|(b\bar{c})\rangle$ quarkonium events and $1.0\times10^4$  $|(b\bar{b})\rangle$ bottomonium events per year can be obtained. Therefore we need to take these higher excited states into consideration for a sound estimation.

\hspace{2cm}

{\bf Acknowledgements}: We are grateful to Xing-Gang Wu for many enlightening discussions.

\end{document}